\documentclass{iopjournal}

\begin{document}

\articletype{Paper} 

\title{Towards Continuous-variable Quantum Neural Networks for Biomedical Imaging}

\author{Daniel Alejandro Lopez$^1$\orcid{0000-0002-3546-6762}, Oscar Montiel$^{1,*}$\orcid{0000-0002-7060-9204}, Oscar Castillo$^2$\orcid{0000-0002-7385-5689} and Miguel Lopez-Montiel$^{3}$\orcid{0000-0001-5367-9801}}

\affil{$^1$Instituto Politécnico Nacional, CITEDI, Tijuana, México}

\affil{$^2$Tijuana Institute of Technology, TecNM, Tijuana, México}

\affil{$^3$CETYS, CETYS Universidad, Tijuana, México}

\affil{$^*$Author to whom any correspondence should be addressed.}

\email{oross@ipn.mx}


\begin{abstract}
Continuous-variable (CV) quantum computing offers a promising framework for scalable quantum machine learning, leveraging optical systems with infinite-dimensional Hilbert spaces. While discrete-variable (DV) quantum neural networks have shown remarkable progress in various computer vision tasks, CV quantum models remain comparatively underexplored. In this work, we present a feasibility study of continuous-variable quantum neural networks (CV-QCNNs) applied to biomedical image classification. Utilizing photonic circuit simulation frameworks, we construct CV quantum circuits composed of Gaussian gates, such as displacement, squeezing, rotation, and beamsplitters to emulate convolutional behavior. Our experiments are conducted on the MedMNIST dataset collection, a set of annotated medical image benchmarks for multiple diagnostic tasks. We evaluate CV-QCNN's performance in terms of classification accuracy, model expressiveness, and resilience to Gaussian noise, comparing against classical CNNs and equivalent DV quantum circuits. This study aims to identify trade-offs between DV and CV paradigms for quantum-enhanced medical imaging. Our results highlight the potential of continuous-variable models and their viability for future computer-aided diagnosis systems.
\end{abstract}

\section{Introduction}\label{sec:introduction}
In recent years, Artificial Intelligence (AI) methods such as Machine Learning (ML) and Deep Learning (DL) have showcased promising potential in the CAD field, demonstrating diagnostic capabilities in diseases such as skin cancer \cite{Esteva2017}, pneumonia \cite{Kermany2018}, and COVID-19 \cite{Wang2020}. Across multiple medical imaging modalities, such as X-rays, Computer Tomography (CT), and Magnetic Resonance Imaging (MRI), AI-based systems have demonstrated the ability to accelerate diagnosis and improve clinical decision-making. However, these methodologies rely heavily on large-scale data processing and computational resources. As the complexity of medical imaging tasks grows, so does the requirement for larger models and massive data processing pipelines \cite{Lecun2015}. Hence, different approaches have been considered to address these issues, such as edge computing, and quantum computing, each leveraging different paradigms and viewpoints \cite{Dunjko2017}. 

Quantum computing (QC) offers a fundamentally different approach to computation, exploiting the principles of superposition, entanglement, and interference to perform operations on exponentially large state spaces \cite{Nielsen2010}. Thanks to the fundamental unit of information, the qubit, both 0 and 1 can be represented simultaneously, allowing parallel computations. This parallelism underlies the  quantum advantage, also known as \textit{quantum speedup}, where certain quantum algorithms outperform their classical counterparts. Examples such as Shor's algorithm for integer factorization \cite{Shor1996}, and Grover's algorithm \cite{Grover1996} for unstructured search, demonstrated exponential or quadratic improvements in runtime compared to classical algorithms. Furthermore, quantum computing has also made a significant impact in various healthcare applications, such as molecular simulation, precision medicine, and drug discovery and development \cite{raihan2023, ullah2024}. As a result, the integration of quantum computing and artificial intelligence has led to the emergence of the field of Quantum Machine Learning (QML), which exploits the properties of both disciplines to enhance classical algorithms and data management \cite{Schuld2014}. In the case of healthcare, improving clinical studies, medical device inspections, and disease diagnosis \cite{ullah2024}.

QML aims to enhance machine learning algorithms by integrating quantum mechanical operations via Parameterized Quantum Circuits (PQCs) to solve tasks such as regression, clustering, or classification \cite{Schuld2018}. A key aspect of QML is data encoding, which allows to leverage quantum properties in computation by transforming classical input data into quantum states for further processing through quantum algorithms. The two main paradigms in this field are:
\begin{itemize}
    \item \textbf{Discrete-Variable (DV) Quantum Computing:} This paradigm uses qubits as information carriers, which correspond to a two-dimensional Hilbert space per unit. This systems are the foundation of many QML models due to their compatibility with modalities of superconducting and ion-trap quantum hardware. Nevertheless, the number of qubits, gate fidelity, and limited representation space remain limiting factors \cite{Schuld2018}.
    \item \textbf{Continuous-Variable (CV) Quantum Computing:} Operates on quantum harmonic oscillators, known as qumodes, which are described by quadrature operators in an infinite-dimensional Hilbert space. CV quantum computing allow the manipulation of continuous information and offers practical implementations using photonic systems \cite{Killoran2018}.
\end{itemize}

Hence, in this work, we explore the use of Gaussian CV quantum computing in biomedical imaging by proposing a small-scale Gaussian CV Variational Quantum Circuit (VQC) that functions as a Quantum Neural Network (QNN), for classification of the BreastMNIST \cite{breastmnist}, OrganAMNIST \cite{organamnist}, and PneumoniaMNIST \cite{pneumoniamnist2018} from the MedMNIST \cite{medmnistv2} dataset. This proposed CV QNN is compared to a proposed DV QNN of similar parameters and quantum gates, as well as a same-scale classical neural network. To assess classification performance, accuracy, recall, precision, and F1 score are computed, as well as the area under the Receiver Operating Characteristic (ROC) and Precision-Recall (PR) curves. Additionally, noise robustness testing, statistical analysis, and Grad-CAM computation are conducted to assess generalization degree, output interpretability, as well as, clinical implementation feasibility. 

Within this framework, we introduce our proposed models and summarize the contributions of this work as follows:
\begin{itemize}
    \item We propose a small-scale Gaussian continuous-variable quantum neural network for binary and multiclass classification on MedMNIST, exploring the underdeveloped field of CV quantum computing on computer-aided diagnosis.
    \item We propose a small-scale discrete-variable quantum counterpart for binary and multiclass classification on MedMNIST.
    \item We evaluate and demonstrate Gaussian noise robustness for the proposed quantum models.
    \item We demonstrate through statistical analysis and hypothesis testing that the proposed quantum models attain comparable image classification to their classical counterpart under this configuration.
    \item We release all code, trained weights, and logs to ensure transparency and replication.
\end{itemize}

This manuscript is structured as follows. Section \ref{sec:introduction} introduces the computer-aided diagnosis field, the potential and role of quantum computing in healthcare, as well as its integration with artificial intelligence in quantum machine learning. Section \ref{sec:relatedWork} reviews recent advancements in discrete-variable and continuous-variable quantum machine learning, with a strong focus on healthcare and computer vision tasks. Section \ref{sec:methodology} details the proposed methodology, including data preparation, the proposed quantum model architectures, and evaluation metrics. Section \ref{sec:results} presents the experimental results by divided into 6 main tests, that outline the attained results for the proposed CV and DV quantum models, as well as their classical counterpart. Section \ref{sec:discussion} goes over the interpretation of the attained classification performance of all models for each of the evaluated datasets. Finally, Section \ref{sec:conclusionAndFutureWork} provides the work's conclusions, emphasizing the significance of the attained results, and reviewing potential directions in the next steps.
\section{Related Work}\label{sec:relatedWork}
The field of Computer-Aided Diagnosis (CAD) has advanced through diverse methodologies, ranging from enhanced data processing, automated data workflows, artificial intelligence and more recently, quantum computing approaches. This review focuses on some of the potential improvements quantum computing presents in healthcare, as well as developments in the quantum machine learning field from the discrete-variable and continuous-variable paradigms.

In \cite{Schuld2018, Schuld2019}, the potential of the implementation of quantum algorithms and quantum data encoding for ML tasks was discussed and demonstrated. Input data can be encoded into quantum states and represented in high-dimensional Hilbert spaces, where data is implicitly defined as kernel functions between data points and can be exploited by kernel-based methods, such as the Support Vector Machine. Following this and further development of the QML field in ML and DL tasks, QML showed promising results in healthcare, as seen in \cite{Sengupta2021}, where clinical prognostic analysis for image classification and segmentation of COVID-19 is accelerated, outperforming conventional deep learning methods by 2.92\%. Similarly, it has been able to predict heart disease through an ensemble model that works as a quantum support vector machine for classification, as demonstrated in \cite{Abdulsalam2022}, where it attained an accuracy of 90.16\% competing with state-of-the-art models. Further research has been conducted on other areas that QML can aid, such as drug response as shown in \cite{Sagingalieva2023}, where a hybrid quantum neural network based on a graph and deep convolutional neural layers is proposed, outperforming classical analogs by 15\% in drug effectiveness prediction. 

In recent years, QML models have shown potential (82.86\% accuracy) in more complex computer vision tasks in the CAD field such as skin lesion classification in dermoscopic images as seen in \cite{Reka2024}, leveraging rotational gates for encoding, as well as classical backbones for feature extraction and preparation for a quantum support vector classifier. Simlarly, favorable results on multiclass classification has been demonstrated, as demonstrated in \cite{Ara2025}, where an 80.96\% diagnosis accuracy was achieved for diabetic retinopathy through a hybrid quantum-classical framework based on ResNet50 and an 8-qubit quantum classifier. Moreover, innovative approaches such as quantum-enhanced dual-backbone architecture as shown in \cite{Marzoug2025} have been developed, achieving a parameter complexity reduction of 29.04\% and 94.44\% of trainable parameters, while still attaining high accuracy (95.80\% and 95.42\% on training and validation sets). 

In contrast, CV QML has not had an equal surge, mainly due to photonic quantum computers still being in development. However, Hilbert space data representation can be extended in CV quantum computing, as data embedded into infinite-dimensional Hilbert spaces via Gaussian gates presents potential to more refined data features and details than classical methods. Because of this, early CV QML research can be traced back to \cite{Lau2017}, where a set of QML subroutines are generalized for infinite-dimensional systems intended for an all-photonic CV quantum computer. Furthermore, general methods for building CV neural networks for CV quantum computers is introduced in \cite{Killoran2018}, where information encoding and nonlinear activation functions are enacted through Gaussian and non Gaussian gates. 

Moreover, machine learning and optimization techniques for quantum photonic circuits were shown in \cite{Moody2019}. Here, a network comprised of several layers of optical gates with variable parameters are optimized applying automatic differentiation, showing the power and versatility of learning how to effectively use short-depth circuits to synthesize single-photons. Additional CV kernel studies are conducted in \cite{Li2022}, where the expressiveness of large Hilbert spaces is explored, introducing quantum kernel encoding methods into CV quantum states through amplitude squeezing and phase manipulation. Realistic implementations of neural networks on photonic quantum computers is proposed in \cite{Ghasemian2023}, where quantum circuits built in CV architecture encode information in spectral amplitude functions of single-photons.

Some of the first implementations of CV QML in medical tasks can be seen in \cite{Kairon2021}, where COVID-19 diagnosis is conducted through CV Quantum Neural Networks (QNNs), comparing its performance with a quantum backpropagation multilayer perceptron is analyzed. Furthermore, the multiclass classification task is tackled in \cite{Choe2022}, where based on the CV architecture proposed in \cite{Killoran2018}, a MNIST classifier is proposed, focusing on the number of encoding qumodes, and using Gaussian and non Gaussian gates for bias addition and nonlinear functions. In recent years, CV QNN development has also been implemented in time-series forecasting as shown in \cite{Anand2024}. In \cite{Anand2024}, a comparison with the DV quantum and classical counterpart is conducted for energy consumption and stock price data forecasting, showing favorable results with lesser number of parameters for the CV model, while also introducing continuous values and nonlinearities, a problem in qubit-based quantum computing.

Taking this into consideration, although quantum machine learning progress on the CV paradigm has been made, focus on medical imaging is still limited. As a result, we propose a CV QNN for biomedical image classification on datasets from MedMNIST \cite{medmnistv2}, assessing through noise robustness tests, statistical analysis, and comparison to its DV QNN and classical model counterparts.
\section{Methodology}\label{sec:methodology}
In this section, the methodology employed to conduct this research is detailed, addressing the data preparation process, the proposed discrete-variable and continuous-variable hybrid quantum models, their classical counterpart, as well as the experiments and evaluation metrics to asses their medical image classification performance.

\subsection{Data Preparation}
Data is fundamental for developing effective and robust models, as the model generalization capabilities rely heavily on the extraction of key features that distinguish the dataset's classes. The data used in this work corresponds to the MedMNIST dataset \cite{medmnistv2}, a large-scale collection of standardized biomedical images which includes twelve 2D datasets, and six 3D datasets. All images are of (28, 28) dimensions and include data annotations for computer vision task purposes, such as binary and multiclass classification, regression, and multi-labeling. The proposed models are evaluated on the 2D grayscale datasets of PneumoniaMNIST \cite{pneumoniamnist2018}, OrganAMNIST \cite{organamnist}, and BreastMNIST \cite{breastmnist} corresponding to the conditions of pneumonia, organ identification, and breast cancer, respectively. Every selected dataset represents a different type of dataset, where PneumoniaMNIST is a binary classification dataset with sufficient training samples, BreastMNIST represents one with a few hundred samples, presenting a more difficult generalization challenge to the proposed models. On the other hand, the OrganAMNIST dataset is an 11 multiclass dataset, which evaluates not only the proposed models' generalization, but also the data dimensionality reduction effectiveness, as feature variance across the dataset is so vast that it cannot be represented with only 4 components per sample.

\begin{figure}[ht]
\centering
\includegraphics[width=0.22\textwidth]{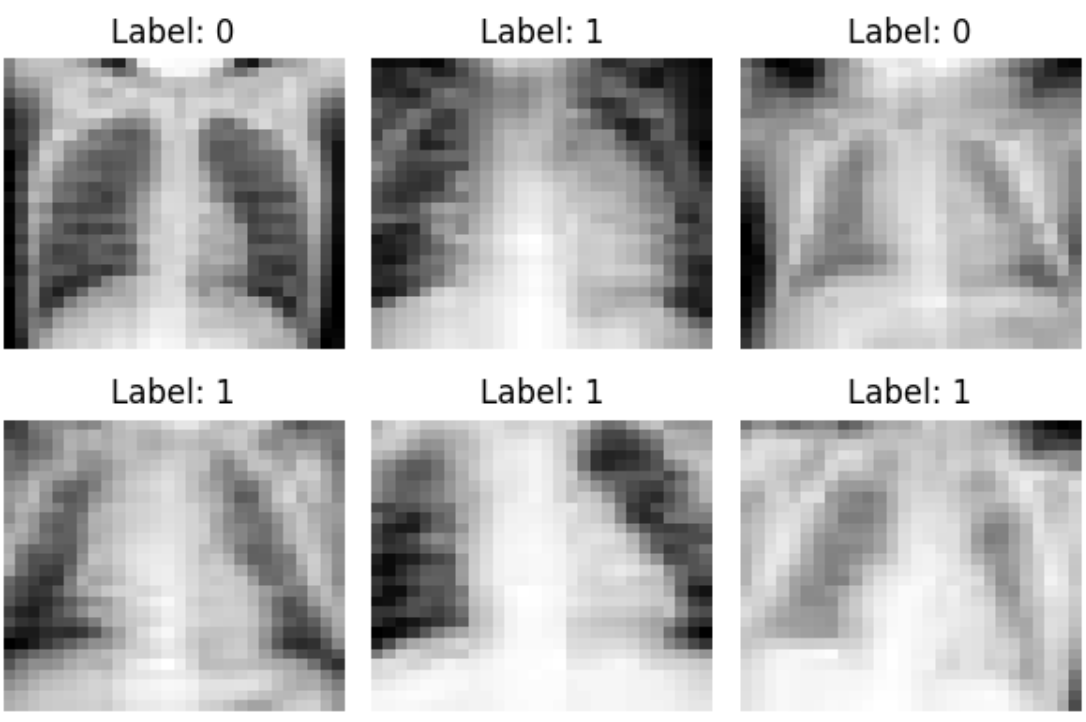}
\hspace{0.3cm}
\includegraphics[width=0.22\textwidth]{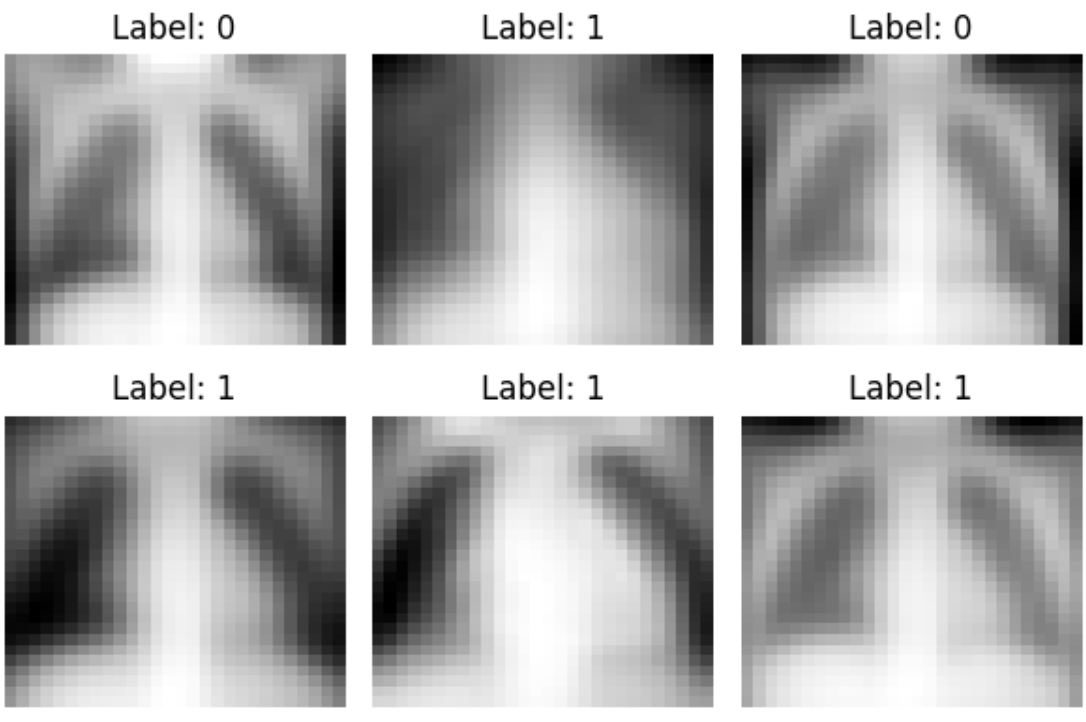}

\vspace{0.3cm}

\includegraphics[width=0.22\textwidth]{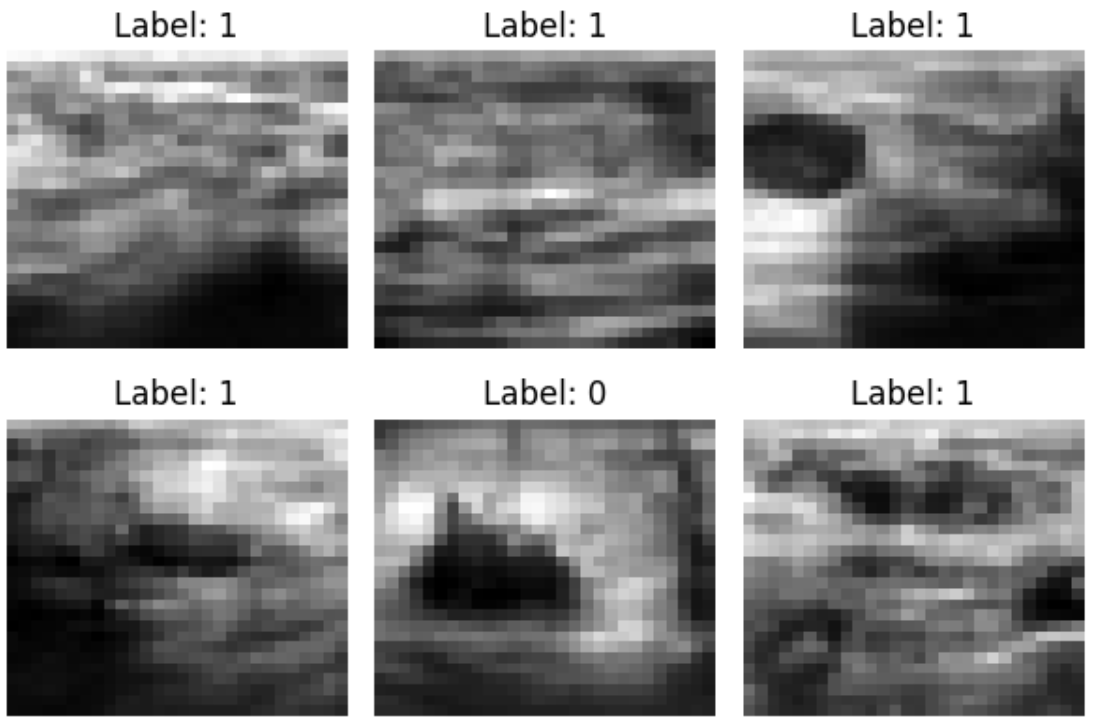}
\hspace{0.3cm}
\includegraphics[width=0.22\textwidth]{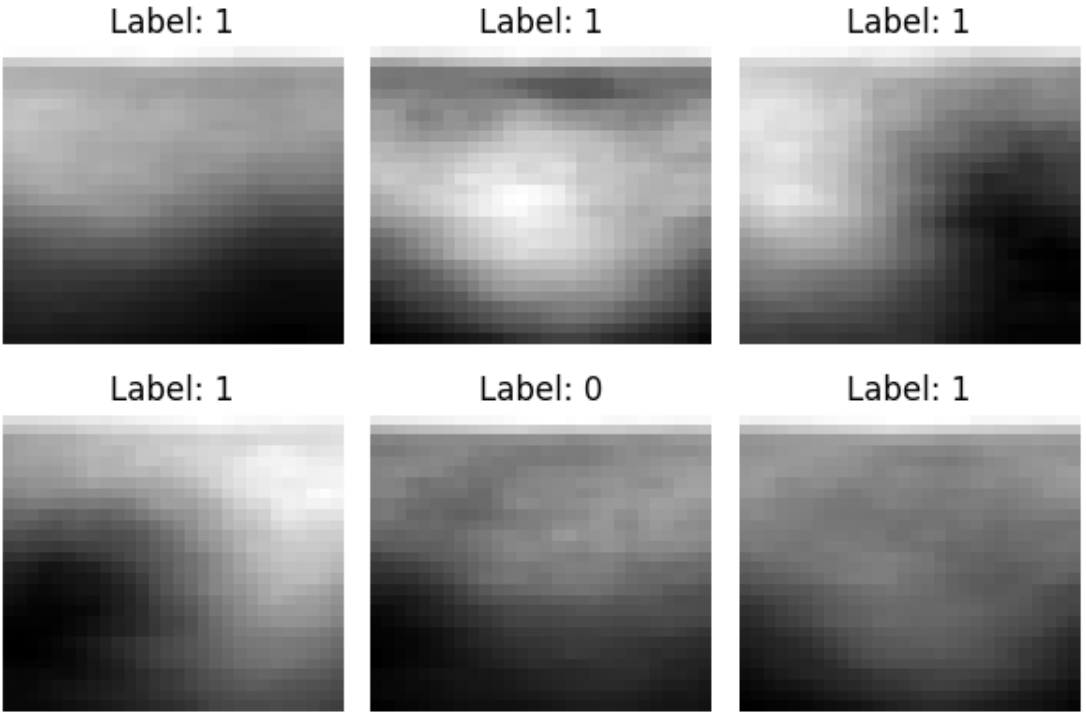}

\vspace{0.3cm}

\includegraphics[width=0.22\textwidth]{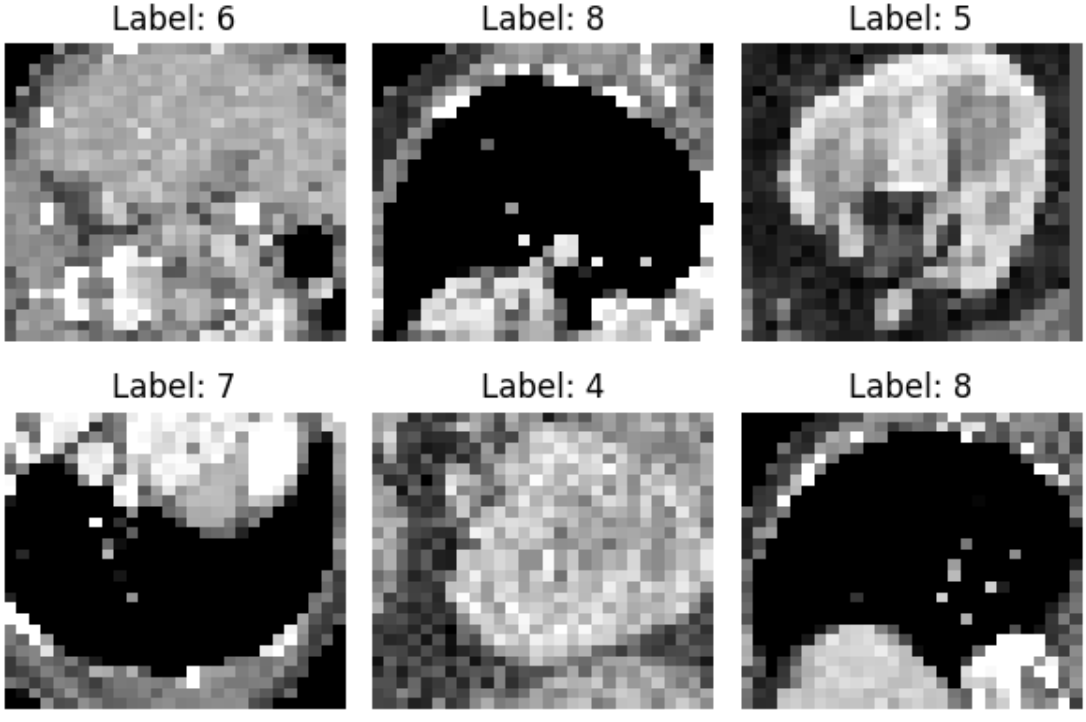}
\hspace{0.3cm}
\includegraphics[width=0.22\textwidth]{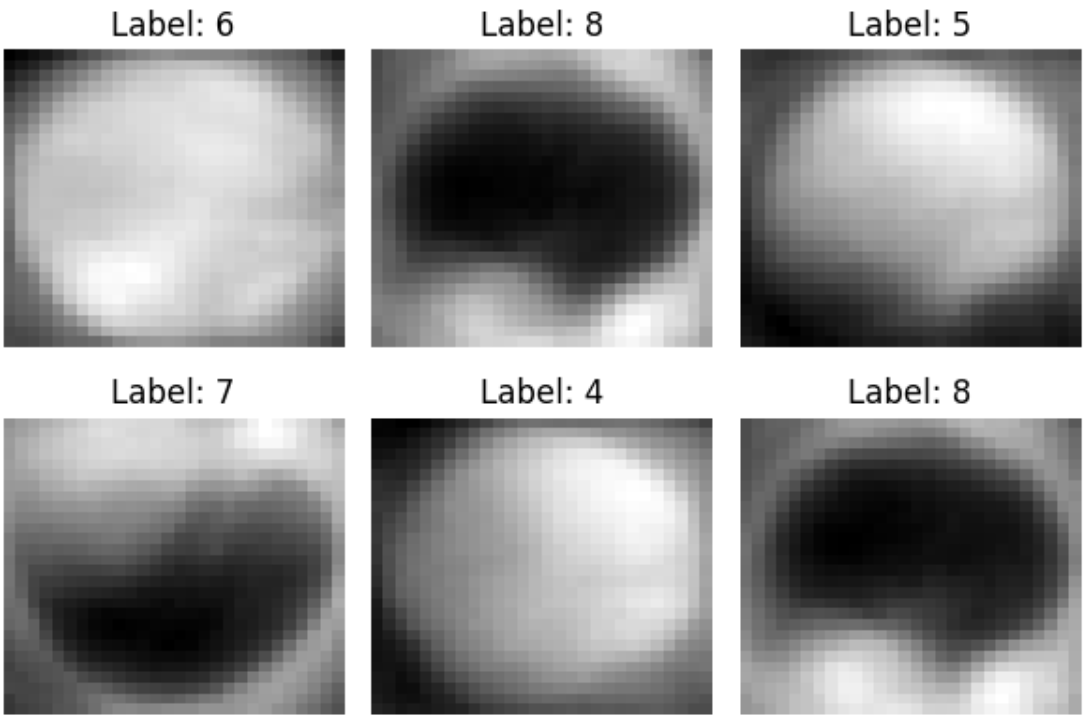}

\caption{
Comparison between original and PCA-reconstructed images for each dataset.  
From top to bottom: PneumoniaMNIST, BreastMNIST, and OrganAMNIST.  
For each dataset, the left image shows an example of the original input, while the right image shows its reconstruction using 4 principal components.
}
\label{fig:pca_reconstruction}
\end{figure}

To accommodate for the small scale of the proposed models due to quantum computing overhead, we employ a Principal Component Analysis (PCA) Encoder \cite{hotelling1933, jolliffe2016} to reduce data dimensionality. Each image, originally sized (28, 28) is flattened into a vector of 784 dimensions. The PCA encoder is then fitted on the training data, this process involves the computation of the mean for each pixel, the capturing of feature correlations through the covariance matrix, and the identification of the directions with maximum variance by eigendecomposition. The top \textit{n} principal components are retained, in this case, projecting each sample into a compact 4-dimensional vector that preserves the most representative patterns. Figure \ref{fig:pca_reconstruction} shows a comparison between the original images and the reconstructed ones from the extracted main four components, where although information variance is maintained to a certain degree, visible differences are apparent.

Furthermore, Table \ref{tab:pca_summary} shows a summary of the data dimension reduction conducted for all the tested datasets, listing the task intended for the dataset, number of samples original dimensions, number of encoded components, and the cumulative variance per dataset by retaining four components. This value represents the intrinsic information content of the dataset preserved in the resulting reduced feature space vector.
\begin{table}[ht]
\scriptsize
\centering
\caption{Dimensionality reduction using PCA across MedMNIST datasets. 
All datasets consist of grayscale images of size $28\times28$ (784 features).}
    \setlength{\tabcolsep}{3.5pt}
    \renewcommand{\arraystretch}{1.75}
\label{tab:pca_summary}
\begin{tabular}{lccccc}
\hline
\textbf{Dataset} & \bf Task & \textbf{Dimensions} & \textbf{No. Samples} & \textbf{PCA} & $\mathbf{\sigma^{2}}$ \\
\hline
Breast      & Binary & ($28,28,1$) & 546  & 4 & $\sim$60\% \\
Organ      & Multiclass & ($28,28,1$) & 10368 & 4 & $\sim$48\% \\
Pneumonia & Binary & ($28,28,1$) & 4708  & 4 & $\sim$60\% \\
\hline
\end{tabular}
\end{table}

\subsection{Continuous-Variable Quantum Neural Network}
The proposed CV quantum circuit leverages the Gaussian formalism implemented in PennyLane Gaussian backend, where each mode of the photonic system corresponds to a quantum harmonic oscillator described in an infinite-dimensional Hilbert space. The circuit employs a combination of parametric Gaussian gates, namely the Displacement $(D)$, Rotation $(R)$, Squeezing $(S)$, and Beamsplitter $(BS)$ gates which together form a universal set for Gaussian transformations. The sections of the CV QNN can be summarized into: data encoding, feature extraction, and data decoding for output assessment and further classical processing. The displacement gate shifts the state in phase space by a complex amplitude $\alpha = r e^{i\phi}$ can be written as:
\begin{equation}
D(\alpha) = \exp\!\big(\alpha \hat{a}^\dagger - \alpha^{\*} \hat{a}\big),
\end{equation}
where $\hat{a}$ and $\hat{a}^\dagger$ denote the annihilation and creation operators, respectively. In the quadrature representation, this transformation corresponds to a translation:
\begin{equation}
\hat{x} \rightarrow \hat{x} + \sqrt{2}\,\mathrm{Re}(\alpha), \qquad
\hat{p} \rightarrow \hat{p} + \sqrt{2}\,\mathrm{Im}(\alpha).
\end{equation}

This gate is employed for data encoding, embedding input features as phase-space displacements for each qumode. Similarly, the rotation gate $R(\phi)$ performs a phase-space rotation by an angle $\phi$, serving as a phase shifter, analogous to the $R_Z(\phi)$ gate in discrete-variable (DV) quantum circuits. This gate can be expressed as:
\begin{equation}
R(\phi) = \exp\!\big(i \phi \hat{a}^\dagger \hat{a}\big),
\end{equation}
which mixes the quadrature operators as:
\begin{equation}
\begin{pmatrix}
\hat{x}' \\[4pt]
\hat{p}'
\end{pmatrix}
=
\begin{pmatrix}
\cos \phi & -\sin \phi \\[4pt]
\sin \phi & \cos \phi
\end{pmatrix}
\begin{pmatrix}
\hat{x} \\[4pt]
\hat{p}
\end{pmatrix}.
\end{equation}

However, gates such as the squeezing gate $S(r)$ focus on parameters exclusive to CV quantum photonic circuits, namely the quadrature, where this gate modifies the uncertainty of a quadrature while increasing it in its conjugate quadrature. Mathematically, this can be written as:
\begin{equation}
S(r) = \exp\!\left[\frac{1}{2}r\!\left(\hat{a}^2 - (\hat{a}^\dagger)^2\right)\right],
\end{equation}
leading to the transformations:
\begin{equation}
\hat{x} \rightarrow e^{-r}\hat{x}, \qquad \hat{p} \rightarrow e^{r}\hat{p}.
\end{equation}
Providing a feature amplification in the latent space, analogous to scaling layers in classical neural networks. In PennyLane’s Gaussian model, the squeezing gate is parameterized by a real squeezing magnitude $r$. For the 2-qubit gate, analogoues to the CNOT gate used in DV quantum circuits, the Beamsplitter gate $BS(\theta, \phi)$ introduces entanglement between two qumodes via a linear optical transformation by mixing the angle $\theta$ and phase $\phi$:
\begin{equation}
BS(\theta, \phi) = 
\exp\!\left[
\theta\!
\left(
e^{i\phi}\hat{a}_1^\dagger \hat{a}_2
- e^{-i\phi}\hat{a}_1 \hat{a}_2^\dagger
\right)
\right],
\end{equation}
producing the following mode transformations:
\begin{equation}
\begin{pmatrix}
\hat{a}_1' \\[4pt]
\hat{a}_2'
\end{pmatrix}
=
\begin{pmatrix}
\cos \theta & -e^{i\phi} \sin \theta \\[4pt]
e^{-i\phi} \sin \theta & \cos \theta
\end{pmatrix}
\begin{pmatrix}
\hat{a}_1 \\[4pt]
\hat{a}_2
\end{pmatrix}.
\end{equation}

\begin{figure}[ht]
\centering
\scalebox{0.75}{
\begin{quantikz}[row sep=0.3cm, column sep=0.5cm]
\lstick{0} & \gate{D} & \gate{R} & \gate{S} & \gate[2]{BS} & \gate{R}  & \gate{S} & \gate[2]{BS} & \meter{} \\
\lstick{1} & \gate{D} & \gate{R} & \gate{S} &              & \gate{R}  & \gate{S} &               & \meter{} \\
\lstick{2} & \gate{D} & \gate{R} & \gate{S} & \gate{R}     & \gate{S}  &          &               & \meter{} \\
\lstick{3} & \gate{D} & \gate{R} & \gate{S} & \gate{R}     & \gate{S}  &          &               & \meter{}
\end{quantikz}
}
\caption{The proposed 4-mode Continuous-Variable (CV) quantum circuit. Each qumode undergoes displacement ($D$) for data encoding, followed by rotational ($R$) and squeezing ($S$) gates for feature extraction. Beamsplitter ($BS$) operations entangle adjacent modes analogously to CNOT gates in DV circuits. Finally, the quadrature expectation values $\langle \hat{X} \rangle$ are measured for data decoding.}
\label{fig:cv-qc}
\end{figure}
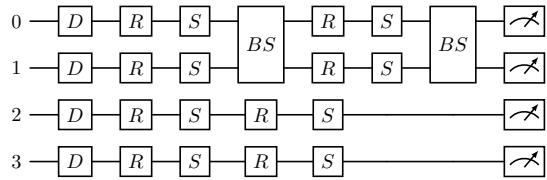

Considering these set of CV Gaussian gates, the architecture of the circuit is defined as:
\begin{itemize}
    \item Local single-mode transformations $\{R(\phi_i), S(r_i)\}$,
    \item Followed by two-mode entangling operations $BS(\theta, \phi)$.
\end{itemize}
Finally, the network output is formed by measuring the expectation values of the position quadratures:
\begin{equation}
\mathbf{y} = \big[\, \langle \hat{X}_1 \rangle, \langle \hat{X}_2 \rangle, \ldots, \langle \hat{X}_n \rangle \,\big].
\end{equation}
These measurements constitute the continuous-variable embedding of the data features which are later passed onto the classical head for the classification task. The diagram of Figure \ref{fig:cv-qc} showcases the proposed CV quantum circuit.

This quantum circuit is implemented as a a neural network layer to define the CV QNN, where the number of trainable parameters depends on the number of classes of the dataset. Due to computational overhead during gradient computation for the quantum gate parameters, architecture is kept relatively small, leading to a small-scale CV QNN comprised of:

\[
\scalebox{0.75}{$
\setlength{\tabcolsep}{3.5pt}
\renewcommand{\arraystretch}{1.75}
\begin{array}{l c c l}
\textbf{Component} & \textbf{Formula} & \textbf{Parameters} & \textbf{Description} \\
\hline
\text{Quantum CV Layer} & 2 \times 4 \times 4 & 32 & (D, R, S, BS) \\
\text{Classical Head} & 4 \times 2 + 2 & 10 & \text{Linear layer mapping} \\
\hline
\textbf{Total} & \text{--} & \textbf{42} & \text{Trainable parameters} \\
\end{array}
$}
\]
The proposed 4-qubit CV QNN comprises a hybrid architecture combining a parameterized quantum circuit and a classical linear layer head for classification. The variational circuit structure is repeated twice, resulting in two layers of learnable quantum operations, and the expectation values of the quantum outputs are measured and passed to the classical head as a fully connected layer with 4 inputs and 2 outputs neurons in the case of the BreastMNIST and PneumoniaMNIST datasets, which when repeated twice and adding their corresponding biases leads to a total of 42 trainable parameters.

\subsection{Discrete-Variable Quantum Neural Network}
To compare the performance of the proposed CV QNN to other quantum computing paradigms, a low-parameter Discrete Variable Quantum Neural Network (DV QNN) is proposed to assess its classification performance for biomedical image diagnosis. Due to the computational constraints of quantum circuit training and gradient calculation, the size and depth of the proposed quantum circuit is kept at 4 qubits, aligning with the dimensionality reduction conducted during data preparation. The same stages of data encoding, feature extraction, and data decoding are present in the proposed DV QNN, the set of quantum gates selected fo this DV quantum circuit corresponds to the closest alignment to the previously defined CV QNN. 

\begin{figure}[ht]
\centering
\scalebox{0.75}{
\begin{quantikz}[row sep=0.3cm, column sep=0.5cm]
\lstick{0} & \gate{R_Y} & \gate{R_Z} & \gate{R_Y} & \ctrl{1} & \gate{R_Z} & \gate{R_Y} & \ctrl{1} & \meter{} \\
\lstick{1} & \gate{R_Y} & \gate{R_Z} & \gate{R_Y} & \targ{}  & \gate{R_Z} & \gate{R_Y} & \targ{}  & \meter{} \\
\lstick{2} & \gate{R_Y} & \gate{R_Z} & \gate{R_Y} & \qw      & \gate{R_Z} & \gate{R_Y} & \qw      & \meter{} \\
\lstick{3} & \gate{R_Y} & \gate{R_Z} & \gate{R_Y} & \qw      & \gate{R_Z} & \gate{R_Y} & \qw      & \meter{}
\end{quantikz}
}
\caption{The proposed 4-qubit DV quantum circuit, comprised of a set of data encoding $R_{y}(\phi)$ gates; a combination of phase $R_{z}(\phi)$ and rotational $R_{y}(\phi)$ gates for feature extraction that emulate rotational and squeezing gates from CV quantum circuits; CNOT gates that entangle data information, analogous to beam splitter gates in CV quantum systems; and data decoding via expectation values measured on the $z$-axis.}
\label{fig:dv-qc}
\end{figure}
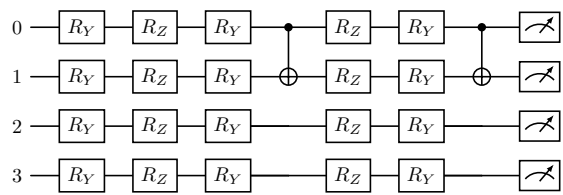

Similarly to the CV QNN, classical data from the PCA-encoded feature vectors is encoded through parameterized y-axis rotations through the $R_{y}(\phi)$ gate, which is represented by:
\begin{equation}
R_{y}(\phi) = e^{-i\phi\sigma_{y}/2} = \begin{bmatrix} \cos(\phi/2) & -\sin(\phi/2) \\
    \sin(\phi/2) & \cos(\phi/2)\end{bmatrix},
\end{equation}
where the $\phi$ angle is the trainable rotation parameter of the circuit. 
The feature extraction stage applies alternating phase rotation gates $R_{z}(\phi)$ and additional $R_{y}(\phi)$ for feature extraction and emulation of rotational and squeezing gates, resembling as closely as possible the proposed CV QNN. The $R_{z}(\phi)$ phase rotation gate can be expressed as:
\begin{equation}
R_{z}(\phi) = e^{-i\phi\sigma_{z}/2} 
= 
\begin{bmatrix}
e^{-i\phi/2} & 0 \\
0 & e^{i\phi/2}
\end{bmatrix}.
\end{equation}
where $\phi$ is the trainable rotation angle around the $z-axis$ of the Bloch sphere, and controls the relative phase between the computational basis states $\ket{0}$ and $\ket{1}$. 
Furthermore, the feature extraction stage requires 2-qubit gates that produce entanglement, which is achieved through the CNOT gates that can be interpreted as analogous to the beam splitter gates of CV quantum systems. The CNOT gate acts on two qubits, where one is determined as the control qubit $q_c$ and the other is the target qubit $q_t$, and the target qubit is flipped only if the control qubit is in state $\ket{1}$. It can be written as as the following:
\begin{equation}
\text{CNOT} 
= |0\rangle\!\langle 0| \otimes I + |1\rangle\!\langle 1| \otimes X =
\begin{bmatrix}
1 & 0 & 0 & 0 \\
0 & 1 & 0 & 0 \\
0 & 0 & 0 & 1 \\
0 & 0 & 1 & 0
\end{bmatrix}.
\end{equation}
Finally, the data decoding stage returns the generated feature maps from the quantum circuit mid-stage to the classical realm for further processing and output interpretation. This is done by measuring along the $z-$axis to obtain the expectation values, which can be written as:
\begin{equation}
    \text{Measurement} = \langle \psi | \sigma_{z} | \psi \rangle,\label{eq:measurement}
\end{equation}
where $\sigma_z$ is the Pauli-Z operator, $\bra{\psi}$ the state before the measurement, and $\ket{\psi}$ the state after measurement. The diagram of Figure \ref{fig:dv-qc} shows the proposed DV quantum circuit, where the first layer of $R_{y}(\phi)$ gates represent data encoding, feature extraction is done from the set of $R_{z}(\phi)$ to the final CNOT gate, and data decoding at the end of the circuit when measuring along the $z-$axis. 

Taking the proposed DV quantum circuit and implementing it as a neural network layer helps define the DV QNN, where the number of trainable parameters changes depending on the number of classes of the dataset, nevertheless, trainable parameters are kept to a minimum to minimize backpropagation computation overhead during training. The final proposed small-scale DV QNN architecture is:
\[
\scalebox{0.75}{$
\setlength{\tabcolsep}{3.5pt}
\renewcommand{\arraystretch}{1.75}
\begin{array}{l c c l}
\textbf{Component} & \textbf{Formula} & \textbf{Parameters} & \textbf{Description} \\
\hline
\text{Quantum DV Layer} & 2 \times 4 \times 4 & 32 & (R_{Y}, R_{Z}, R_{Y}, R_{Z}) \\
\text{Classical Head} & 4 \times 2 + 2 & 10 & \text{Linear layer mapping} \\
\hline
\textbf{Total} & \text{--} & \textbf{42} & \text{Trainable parameters} \\
\end{array}
$}
\]
The proposed 4-qubit DV QNN comprises a hybrid architecture combining a parameterized quantum circuit and a classical linear layer head for classification. The variational circuit structure is repeated twice, resulting in two layers of learnable quantum operations, and the expectation values of the quantum outputs are measured and passed to the classical head as a fully connected layer with 4 inputs and 2 outputs neurons in the case of the BreastMNIST and PneumoniaMNIST datasets, which when repeated twice and adding their corresponding biases leads to a total of 42 trainable parameters.

\subsection{Classification Metrics}
To evaluate classification performance of the proposed models the metrics of accuracy (ACC), precision (P), recall (R), and F1 score (F1) are computed. These metrics correspond to the predicted values of every sample assessed defined as True Positive (TP), True Negative (TN), False Positive (FP), and False Negative (FN). The correctly predicted negative and positive samples are labeled as TP and TN, respectively. On the other hand, incorrectly classified samples for the positive and negative classes are labeled as FP and FN \cite{grandini2020}. To represent the model predictions visually, a confusion matrix is used, which is comprised of cells that show the number of predicted values for each class of the dataset. Figure \ref{fig:confusion_matrix} displays a multiclass confusion matrix, used for problems with two or more classes, in this work, two of the used datasets are for the binary classification task, while the remaining one is intended for multiclass classification. In this diagram, the correctly predicted values are shown in the diagonal, the cell of the evaluated class represents the TP values, while the remaining cells among the diagonal correspond to the TN values, and the rest are FN and FP values horizontally and vertically, respectively \cite{grandini2020}. 

\begin{figure}[ht]
\centering
\scriptsize
\begin{tikzpicture}[
    square/.style={minimum size=1.2cm, draw=black, thick, anchor=center},
]

\definecolor{tpcolor}{RGB}{13,8,135}    
\definecolor{fncolor}{RGB}{240,120,100}  
\definecolor{fpcolor}{RGB}{253,231,100}  
\definecolor{tncolor}{RGB}{84,60,190}    

\matrix (m) [matrix of nodes,
             nodes={square},
             column sep=-\pgflinewidth, row sep=-\pgflinewidth] 
{
 |[fill=tncolor, text=white]| TN & |[fill=fpcolor, text=black]| FP & |[fill=tncolor, text=white]| TN & |[fill=tncolor, text=white]| TN \\
 |[fill=fncolor, text=white]| FN & |[fill=tpcolor, text=white]| TP & |[fill=fncolor, text=white]| FN & |[fill=fncolor, text=white]| FN \\
 |[fill=tncolor, text=white]| TN & |[fill=fpcolor, text=black]| FP & |[fill=tncolor, text=white]| TN & |[fill=tncolor, text=white]| TN \\
 |[fill=tncolor, text=white]| TN & |[fill=fpcolor, text=black]| FP & |[fill=tncolor, text=white]| TN & |[fill=tncolor, text=white]| TN \\
};

\node[above=0.2cm of m.north] {Predicted Class};
\node[rotate=90, left=0.7cm of m.west] {True Class};

\node[below=0.2cm of m-4-1] {Class A};
\node[below=0.2cm of m-4-2] {Class B};
\node[below=0.2cm of m-4-3] {Class C};
\node[below=0.2cm of m-4-4] {Class D};

\node[left=0.2cm of m-1-1, rotate=90] {Class A};
\node[left=0.2cm of m-2-1, rotate=90] {Class B};
\node[left=0.2cm of m-3-1, rotate=90] {Class C};
\node[left=0.2cm of m-4-1, rotate=90] {Class D};

\end{tikzpicture}

\caption{Example of how TP, FP, FN, and TN are defined for a given class (here \textbf{Class B}) in a multiclass confusion matrix. The diagonal cell for Class B corresponds to TP, the rest of row B are FN, the rest of column B are FP, and all other cells are TN.}
\label{fig:confusion_matrix}
\end{figure}

After model predictions are finished, classification metrics may be calculated to assess its performance. The accuracy (ACC) represents model performance across all samples in the dataset, considering correctly predicted positive and negative entries. Moreover, precision (P) and recall (R) represent the model's ability to correctly predict positive samples, while the F1 score (F1) is the combination of P and R and showing its generalization balance, these classification metrics are computed through the prediction values as:
\begin{eqnarray}
    P &=& \frac{TP}{TP + FP}, \nonumber\\
    R &=& \frac{TP}{TP + FN}, \nonumber\\
    ACC &=& \frac{TP + TN}{TP + TN + FP + FN}, \nonumber\\
    F1 &=& 2\left(\frac{P \times R}{P + R}\right).
    \label{eq:multiclassMetricsClass}
\end{eqnarray}

Similarly, the Area Under the Receiver Operating Characteristic Curve (AUROC) is employed to evaluate performance on the test set. The ROC curve illustrates classification performance across all possible thresholds, in a way, representing a set of confusion matrices for each evaluated point \cite{goadrich006}. To compute the ROC curve, the True Positive Rate (TPR) and False Positive Rate must be calculated first. These metrics represent the rate of correctly and incorrectly predicted positive and negative samples, respectively, and they are obtained through the predicted values as:
    \begin{eqnarray}
    FPR &=& \frac{FP}{FP + TN}, \nonumber\\
    TPR &=& \frac{TP}{TP + FN}.
    \label{eq:roc_metrics}
    \end{eqnarray}

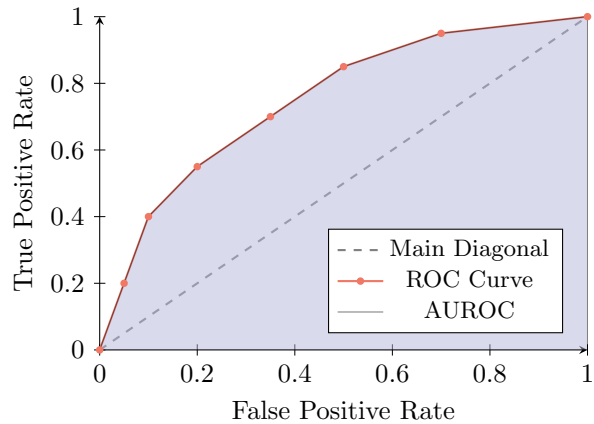
\begin{figure}[ht]
\centering
\begin{tikzpicture}

\definecolor{tpcolor}{RGB}{13,8,135}    
\definecolor{fncolor}{RGB}{240,120,100}  

\begin{axis}[
    width=8cm, height=6cm,
    xlabel={False Positive Rate},
    ylabel={True Positive Rate},
    xmin=0, xmax=1,
    ymin=0, ymax=1,
    axis line style={->},
    axis lines=left,
    grid=none,
    legend style={at={(0.95,0.05)},anchor=south east, font=\small},
    tick style={black},
    xtick={0,0.2,0.4,0.6,0.8,1},
    ytick={0,0.2,0.4,0.6,0.8,1}
]

\addplot[dashed, thick, gray] coordinates {(0,0) (1,1)};
\addlegendentry{Main Diagonal}

\addplot[fncolor, thick, mark=*, mark size=1pt] 
coordinates {
 (0,0) (0.05,0.2) (0.1,0.4) (0.2,0.55) (0.35,0.7) (0.5,0.85) (0.7,0.95) (1,1)
};
\addlegendentry{ROC Curve}

\addplot[fill=tpcolor!30, opacity=0.5] 
coordinates {
 (0,0) (0.05,0.2) (0.1,0.4) (0.2,0.55) (0.35,0.7) (0.5,0.85) (0.7,0.95) (1,1) (1,0) (0,0)
};

\addlegendentry{AUROC}

\end{axis}
\end{tikzpicture}

\caption{Receiver Operating Characteristic (ROC) curve. The red line indicates the ROC curve with threshold points, the dashed gray line represents the main diagonal (random performance), and the shaded blue area corresponds to the area under the ROC curve (AUROC).}
\label{fig:auroc-example}
\end{figure}

Once the FPR and TPR values are obtained, the ROC curve is formulated. The plot in Figure \ref{fig:auroc-example} showcases an example of a binary ROC curve obtained by computing model classification over a set of different thresholds, and showcasing the overall performance. Every threshold point on the curve represents the minimum probability value required for a sample to be classified as positive; otherwise, it is classified as negative.
After every determined threshold is evaluated, the ROC can be expressed mathematically as:
\begin{equation}
    ROC(\sigma) = (FPR(\sigma), TPR(\sigma)),
\end{equation}
FPR and TPR are evaluated over each threshold $\sigma$ \cite{villardon2022}. One more important metric that encapsulates classification performance is the area under the computed ROC curve (AUROC), as shown in the shadowed region of Figure \ref{fig:auroc-example}. The AUROC for a given threshold represented by $\sigma$ is computed as:
\begin{equation}
    AUROC(\sigma) = \int^{b}_{a} TPR(\sigma)d(FPR(\sigma).
\end{equation}
Furthermore, when working with multiclass problems, the AUROC curve configuration used is ``One vs. Rest'' (OvR), which compares every evaluated class against all other, essentially treating one as the positive class and the remainder as negative. The AUROC OvR metric is computed as a function $f = (f^{1}, \ldots, f^{N_c})$, giving an AUC score for each $f^{i}$, where the $i$-th class is positive, and all other classes $j$ are negative \cite{villardon2022}. This is mathematically expressed as:
\begin{equation}
    AUROC^{OvR}(f) = \frac{1}{N_C}\sum^{N_C}_{i=1} AUROC_{i \vert \neg i}(f^{i}). \label{eq:OvR}
\end{equation}

\begin{figure}[ht]
\centering
\begin{tikzpicture}

\definecolor{tpcolor}{RGB}{13,8,135}    
\definecolor{tncolor}{RGB}{84,60,190}    

\begin{axis}[
    width=8cm, height=6cm,
    xlabel={Recall},
    ylabel={Precision},
    xmin=0, xmax=1,
    ymin=0, ymax=1,
    axis line style={->},
    axis lines=left,
    grid=none,
    legend style={at={(0.95,0.05)},anchor=south east, font=\small},
    tick style={black},
    xtick={0,0.2,0.4,0.6,0.8,1},
    ytick={0,0.2,0.4,0.6,0.8,1}
]

\addplot[dashed, thick, gray] coordinates {(0,0.2) (1,0.2)};
\addlegendentry{Baseline}

\addplot[tpcolor, thick, mark=*, mark size=1pt] 
coordinates {
 (0.0,1.0) (0.1,0.9) (0.2,0.85) (0.4,0.75) (0.6,0.55) (0.8,0.40) (1.0,0.25)
};
\addlegendentry{PR Curve}

\addplot[fill=tncolor!30, opacity=0.5] 
coordinates {
 (0.0,0.0) (0.0,1.0) (0.1,0.9) (0.2,0.85) (0.4,0.75) (0.6,0.55) (0.8,0.40) (1.0,0.25) (1.0,0.0) (0.0,0.0)
};
\addlegendentry{AUPRC}

\end{axis}
\end{tikzpicture}

\caption{Example Precision–Recall (PR) curve. The blue line indicates the PR curve, the dashed gray line marks the baseline (positive class prevalence, here $0.2$), and the shaded area corresponds to the AUPRC.}
\label{fig:prcurve}
\end{figure}
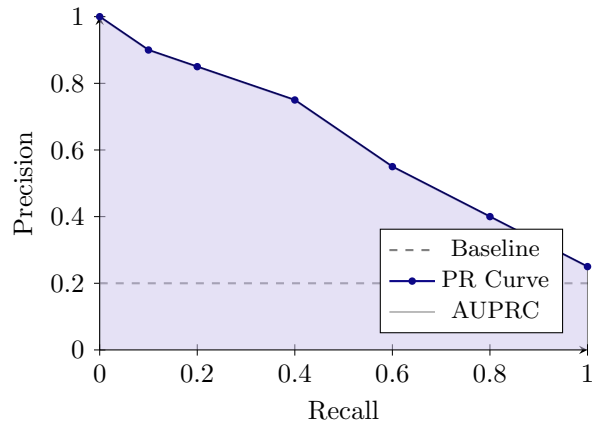

In addition to computing the AUROC curves, we evaluate model performance using Precision-Recall (PR) curves. Similarly to the AUROC curve, this plot illustrates the trade-off between the P and R metrics as the decision threshold varies, aiding in model performance evaluation for imbalanced datasets. The PR curves highlight how well a model balances sensitivity and precision for minority class prediction \cite{goadrich006}. For a binary classification problem, the AUPRC is defined as the area under the precision-recall curve as:
\begin{eqnarray}
\mathrm{AUPRC} = \int_0^1 P(R) \, dR \label{eq:auprc_int} \approx \sum_{i=1}^{n-1} \left( R_{i+1} - R_i \right) P_{i+1} \nonumber,    
\end{eqnarray}
where \(P(R)\) denotes precision as a function of recall, and the discrete approximation follows the trapezoidal integration. For multiclass classification, we apodt the OvR strategy from the AUROC, computing the AUPRC independently for each class \(k\) and averaging across classes. The equation would look be written as:
\begin{eqnarray}
\mathrm{AUPRC}_{\text{OvR}} = \frac{1}{K} \sum_{k=1}^{K} \int_0^1 P_k(R_k) \, dR_k. \label{eq:auprc_ovr}    
\end{eqnarray}
The diagram in Figure \ref{fig:prcurve} shows an example of a precision-recall curve for binary classification, as well as the area under it to compute model trade-off between sensitivity and precision.

Finally, to further assess the proposed models' clinical viability, we plot and evaluate the predictions over the test set using the Gradient-weighted Class Activation Mapping (Grad-CAM) \cite{selvaraju2017gradcam} tool to analyze the models' predictive decision over the highlighted regions of interest across the different evaluated datasets. The Grad-CAM generates a visual heatmap that emphasizes the region of the image that contributed the most to the model's decision. This information allows for a more comprehensive understanding behind the model's predictions, aiding in diagnostic interpretation.

\section{Experiments and Results}\label{sec:results}
This section presents four experiments that evaluate the performance, generalization, and robustness of the proposed continuous variable quantum neural network, discrete variable quantum neural network, as well as a comparison with their classical counterpart. Using the PneumoniaMNIST, BreastMNIST, and OrganAMNIST datasets, we assess classification performance, robustness, and the comprehensibility of their decision heatmap.
Model generalization is examined by conducting threefold cross-validation during training, and computing the standard deviation and mean for every classification metric. Additionally, statistical analysis is conducted through a Friedman and pairwise Wilcoxon tests to determine statistical significant difference in classification performance between the proposed quantum models, as well as their classical counterpart. To ensure reproducibility, all experiments used fixed random seeds for data encoding, shuffling and weight initialization \cite{pineau2021reproducibility}. Training was performed on an Ubuntu 24.04 system with an AMD Ryzen Threadripper 1920X, Pytorch 2.7.0, and Torchvision 0.22.0. For the implementation of the quantum neural networks, the PennyLane 0.29.1 framework was used for Strawberry fields plugin compatibility for comparison purposes. All experiments are run on CPU due to the limitations of the used PennyLane gaussian gates. The chosen hyperparameters are a batch size of 32, a learning rate of $1\times10^{-3}$, the Adam optimizer \cite{kingma2014adam}, and a total of 50 epochs of training for all datasets.

\begin{figure}[ht]
\centering
\includegraphics[width=0.28\textwidth]{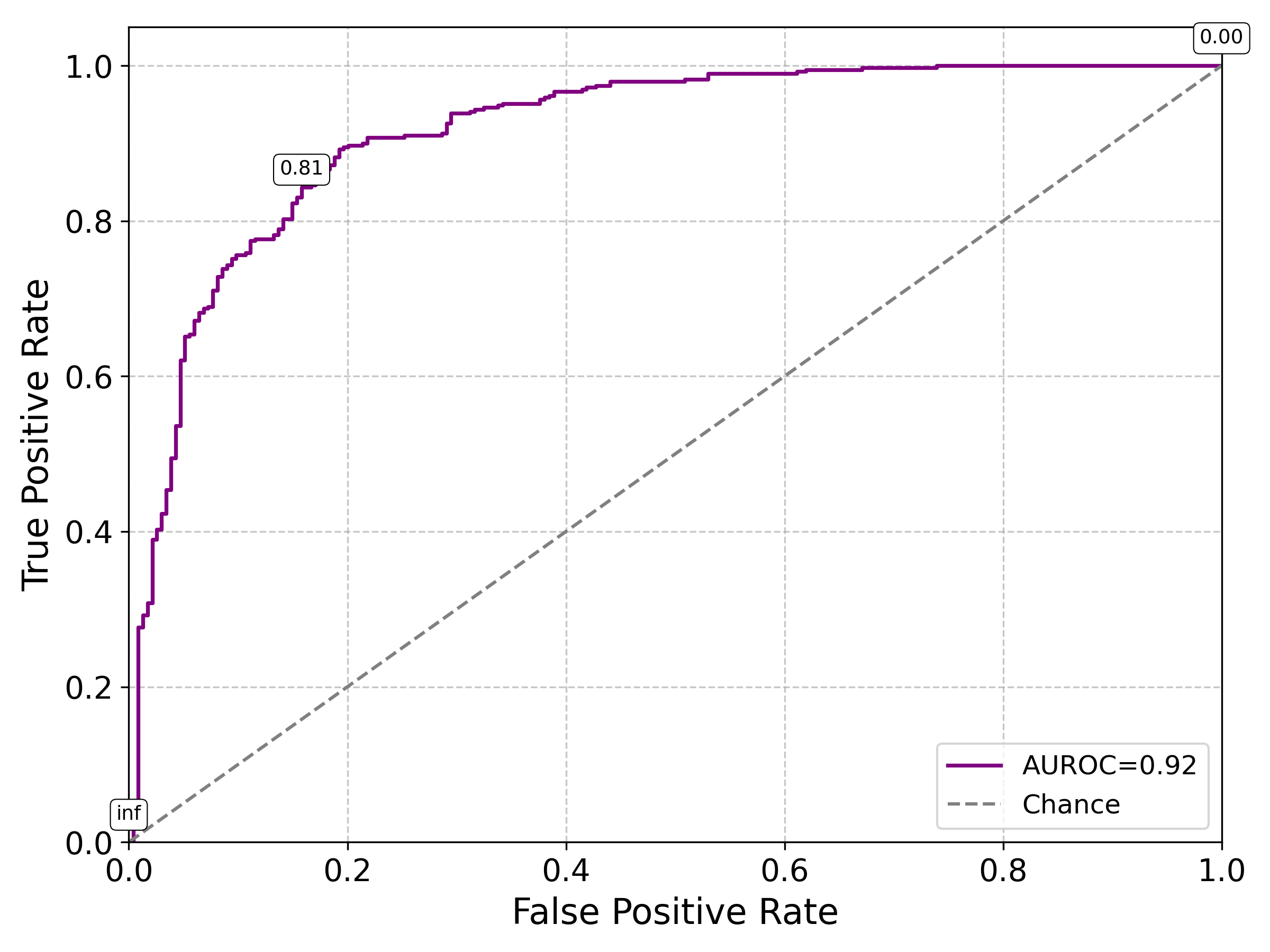}
\hspace{0.4cm}
\includegraphics[width=0.28\textwidth]{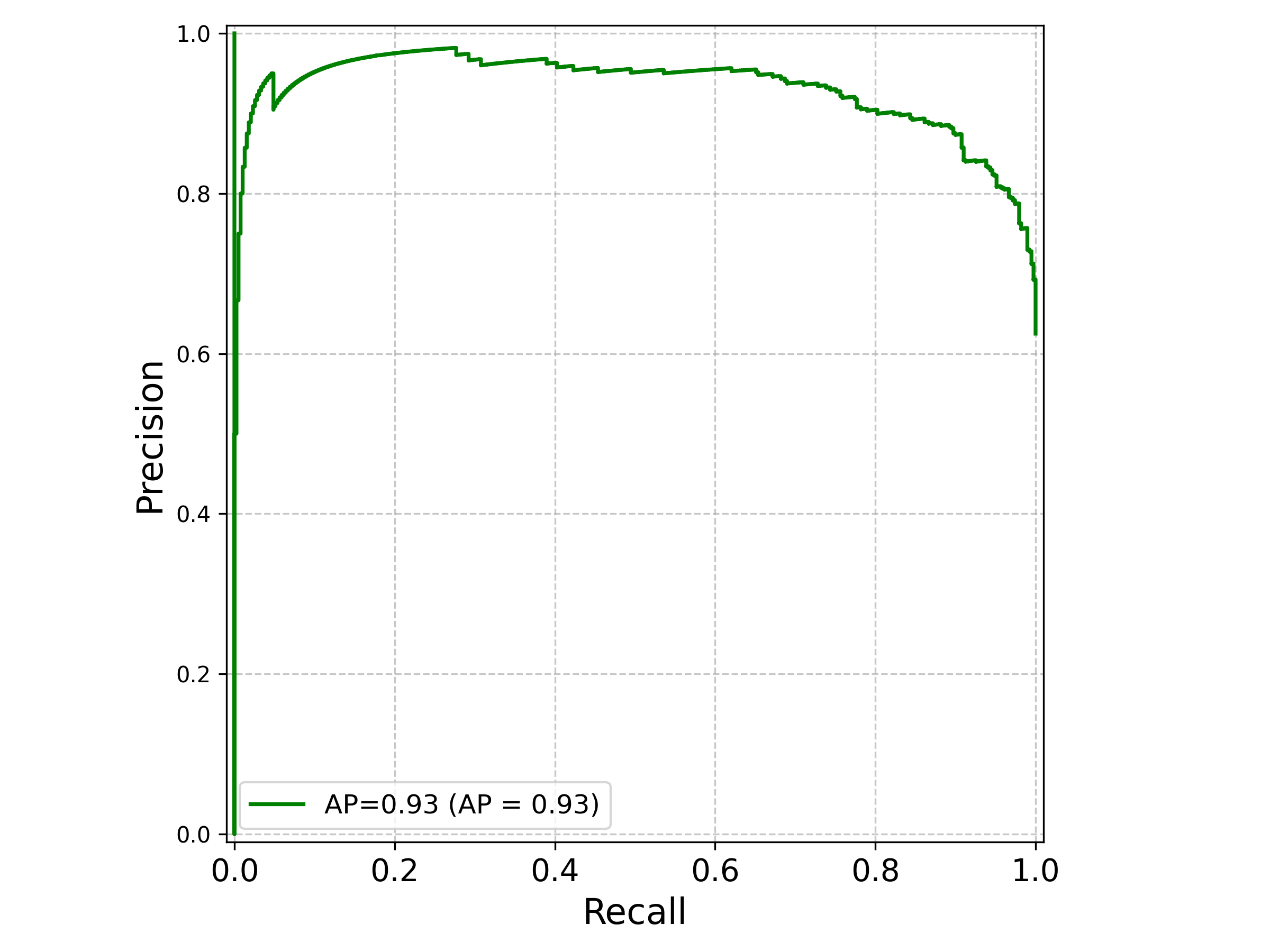}
\hspace{0.4cm}
\includegraphics[width=0.28\textwidth]{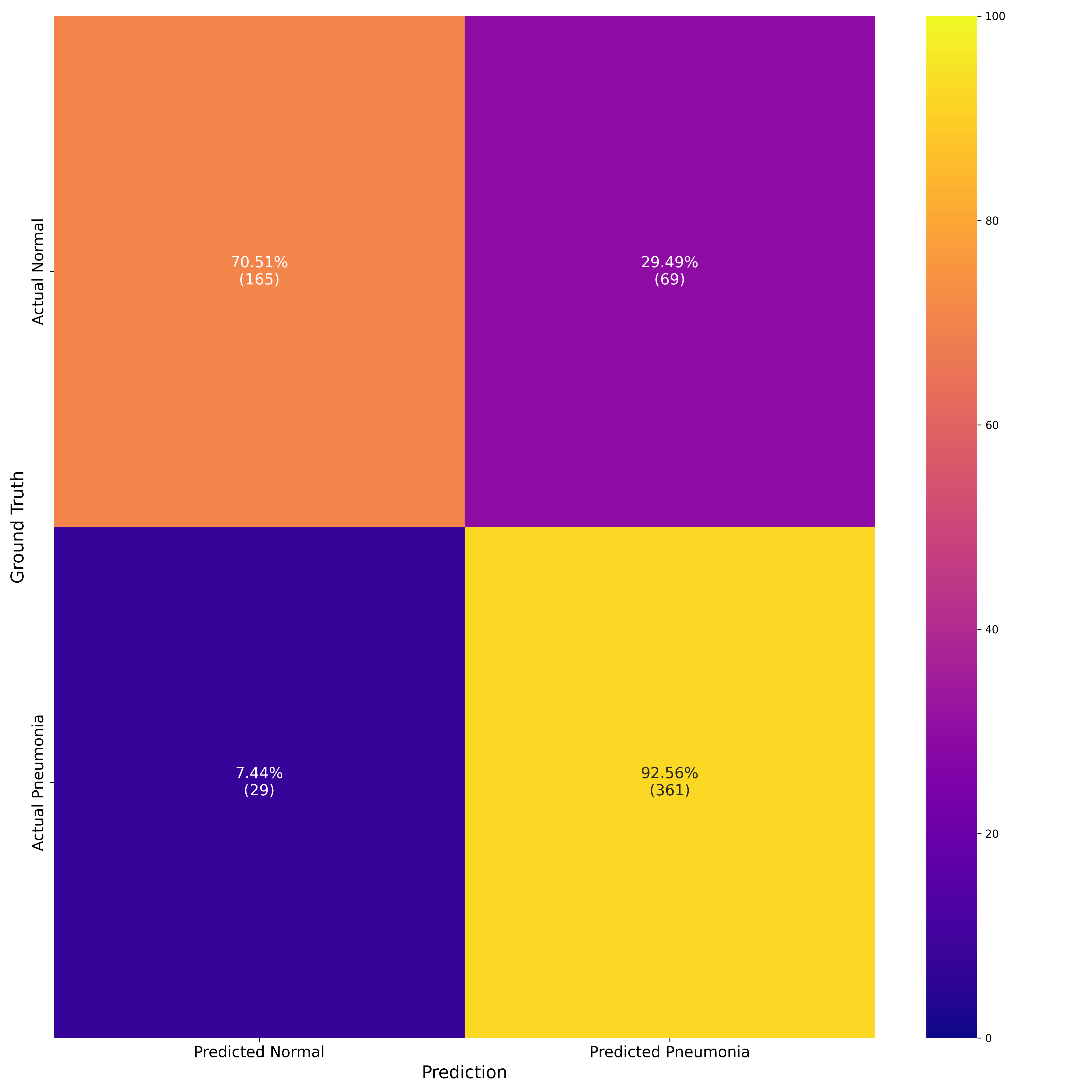}

\caption{
Performance evaluation of the proposed continuous-variable quantum neural network (CV-QNN) on the PneumoniaMNIST dataset.  
From left to right: (a) area under the receiver operating characteristic (AUROC), (b) precision–recall (PR) curve, and (c) confusion matrix.
}
\label{fig:cv-qnn-performance-pneumonia}
\end{figure}

\textbf{Experiment 1: Classification performance on PneumoniaMNIST dataset}\\
The proposed CV quantum neural network is trained through threefold cross-validation with the specified hyperparameters over 50 epochs, obtaining an average accuracy, precision, recall, and F1 score of 89.01\%, 89.01\%, 88.77\%, and 88.81\% across the three folds on the training set, respectively. Similarly, the attained classification performance on the validation set remains consistent across all folds, reaching a mean accuracy of 89.67\%, recall of 89.67\%, precision 89.46\%, and F1 score of 89.67\%, demonstrating its capability of extracting meaningful features through the proposed CV quantum circuit. The training results are further corroborated when evaluating the model's best fold on the test set, where it obtained an accuracy and recall of 84.29\%, precision of 84.37\%, and F1 score of 84.29\%. To further evaluate classification performance, the AUROC curve, PR curve, and confusion matrix on the test set are plotted. The set of plots of Figure \ref{fig:cv-qnn-performance-pneumonia} illustrates the proposed CV quantum neural network's performance at various thresholds to prove robustness. The CV quantum neural network achieved an AUROC of 92\%, and an AUPRC of 93\%; moreover, the plotted confusion matrix shows the total of predicted values and the percentage they represent on the entire dataset. Here, the CV quantum model attained a competitive 70.51\% TN values, which represent normal chest x-rays, and a relatively high 92.56\% TP values, representing correctly identified pneumonia samples. 

\begin{figure}[ht]
    \centering
    \includegraphics[width=0.28\linewidth]{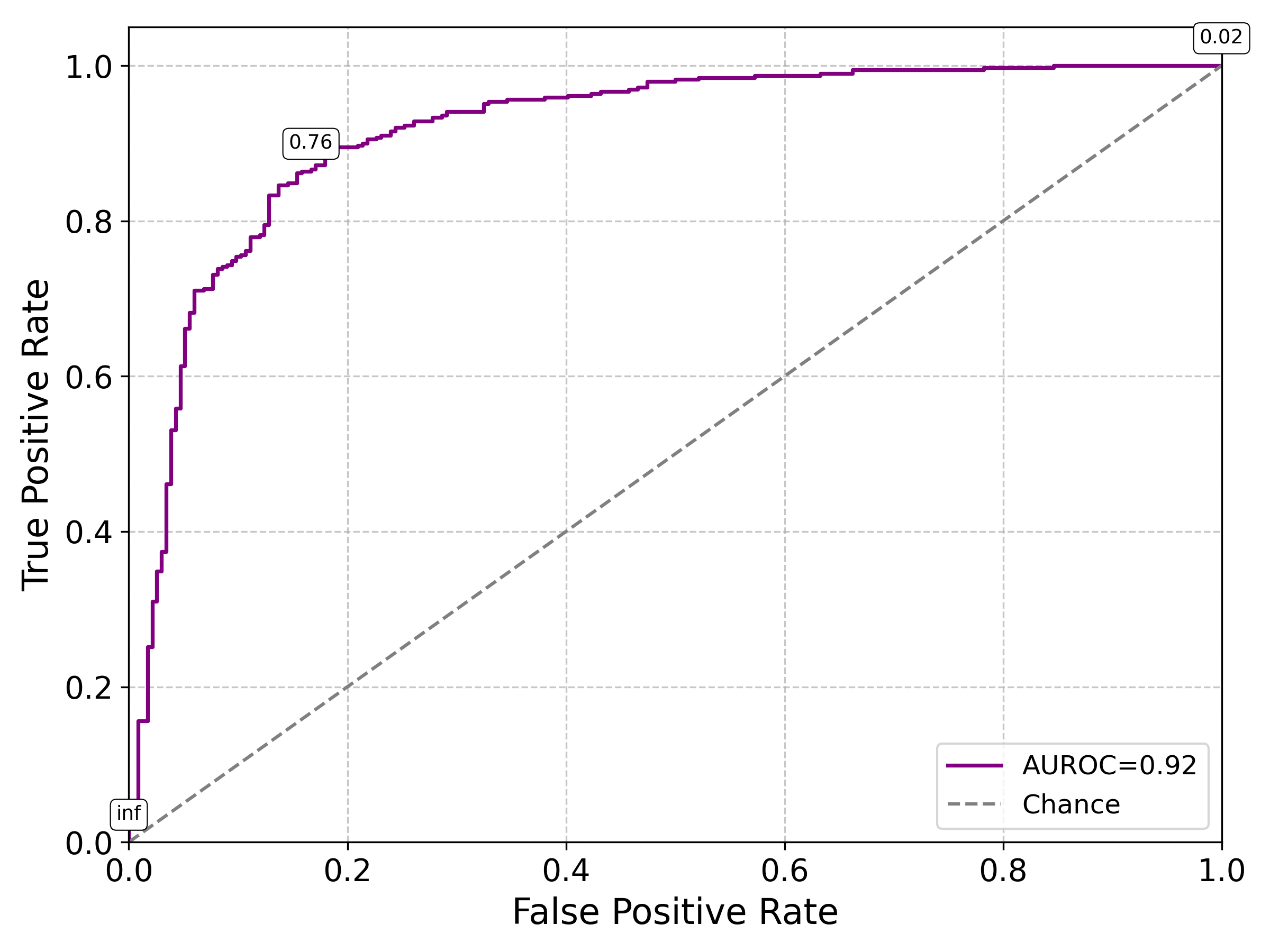}
    \hspace{0.4cm}
    \includegraphics[width=0.28\linewidth]{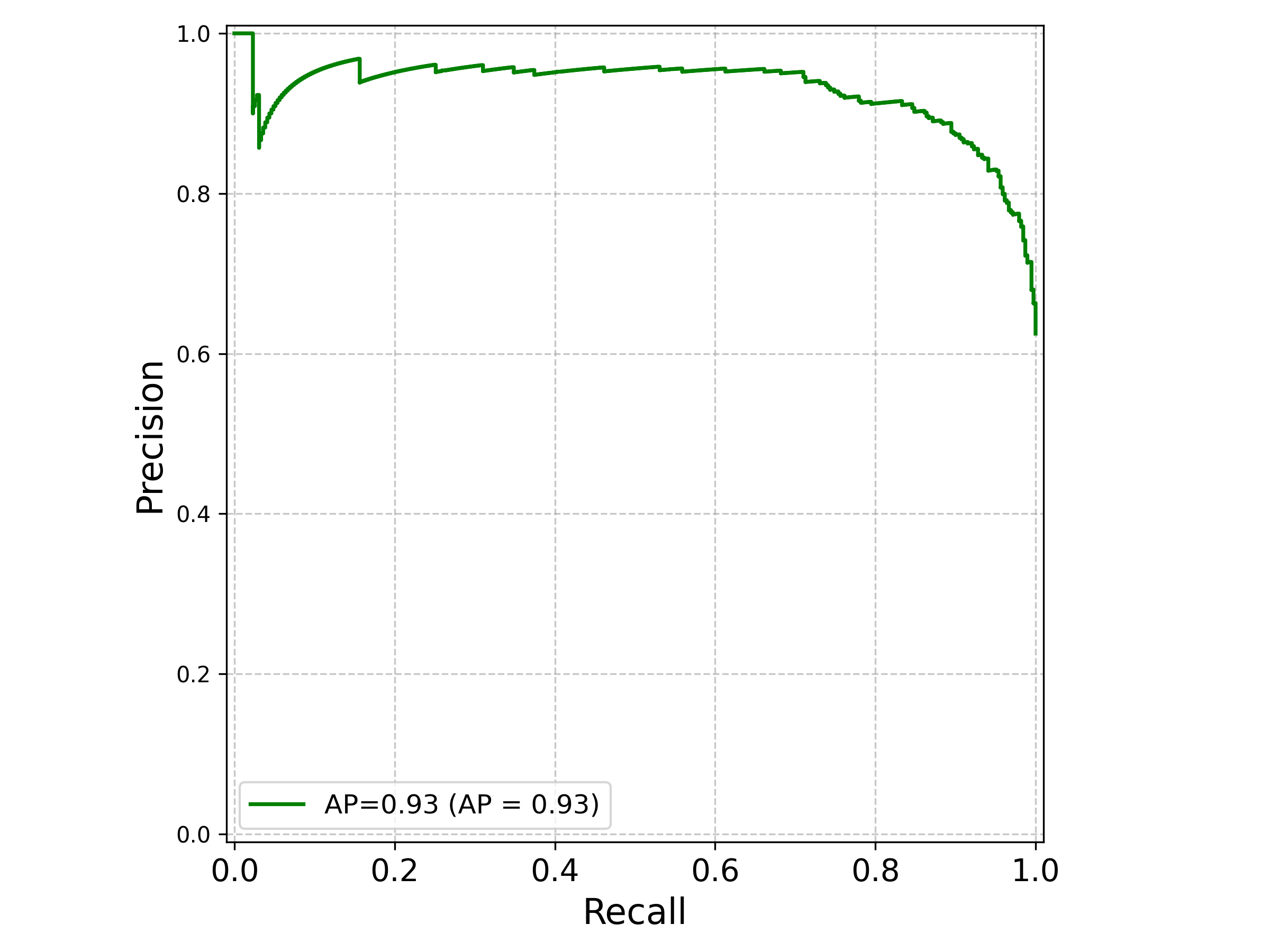}
    \hspace{0.4cm}
    \includegraphics[width=0.28\linewidth]{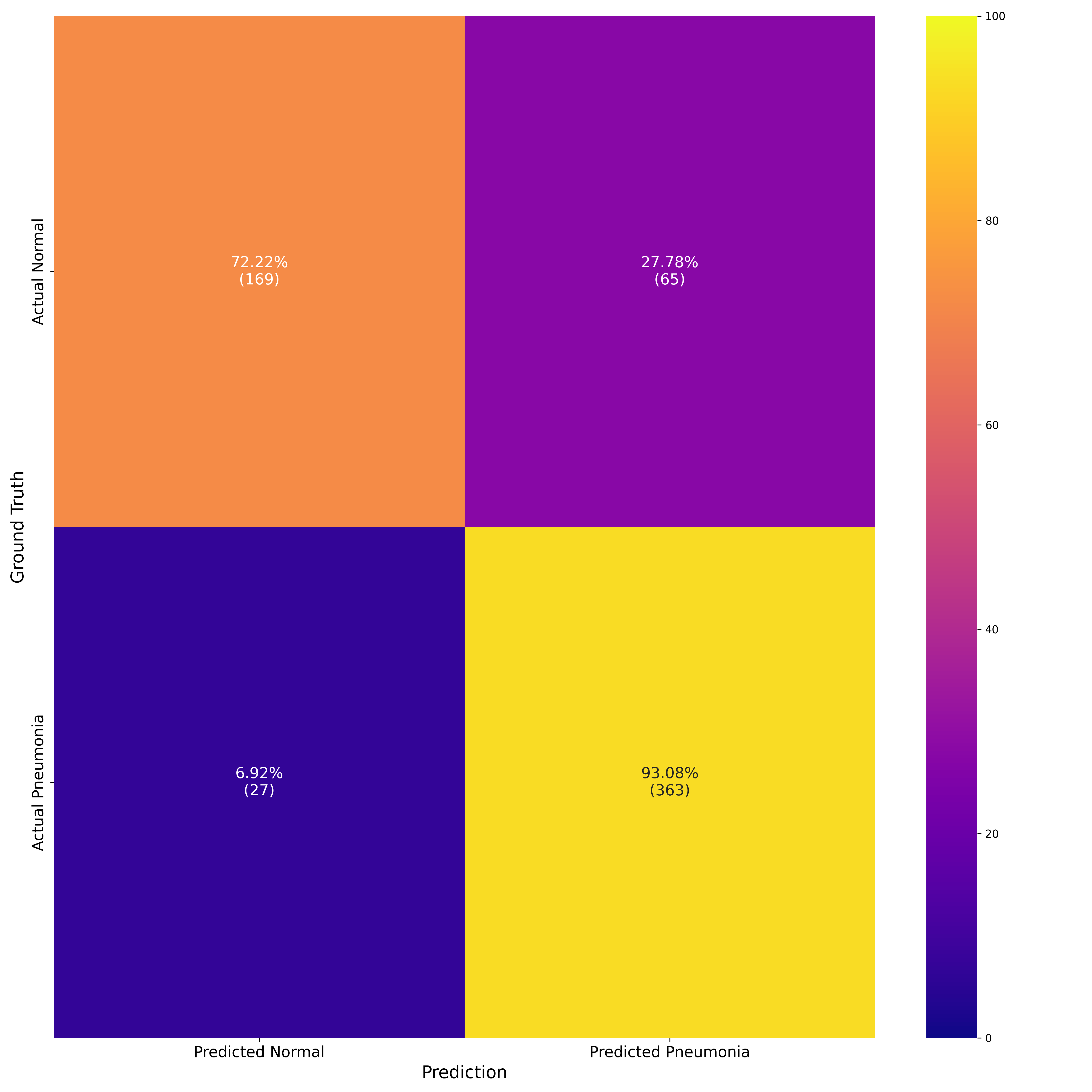}
    
    \caption{Performance evaluation of discrete variable neural network on PneumoniaMNIST dataset: (a) AUROC, (b) precision-recall (PR) curve, (c) confusion matrix.}
    \label{fig:dv-qnn-performance-pneumonia}
\end{figure}

The proposed DV quantum neural network is trained under the same hyperparameters as the CV quantum model for objective comparison. For the training set, the DV model showcased a slight advantage over its CV counterpart, achieving an average of 90.35\% for accuracy and recall, as well as a 90.27\% precision and F1 score. Likewise, the results obtained for the validation set average a 90\% for all classification metrics, exhibiting a small edge when compared to the proposed CV quantum neural network. Additional classification assessment is also conducted on the proposed DV quantum neural network, where it achieved a test accuracy and recall of 85.26\%, a precision of 85.34\%, and a F1 score of 85.26\%. The test set results hold up to what is displayed in the set of plots of Figure \ref{fig:dv-qnn-performance-pneumonia}, where the rate of correctly classified samples for the negative and positive classes on the confusion matrix is 72.22\% and 93.08\%, higher when compared to what the proposed CV quantum model achieved. Similarly, the achieved area under the ROC and PR curves is 92\% and 93\%, respectively, demonstrating comparable performance between both models once implemented on unseen data.

\begin{figure}[ht]
    \centering
    \includegraphics[width=0.28\linewidth]{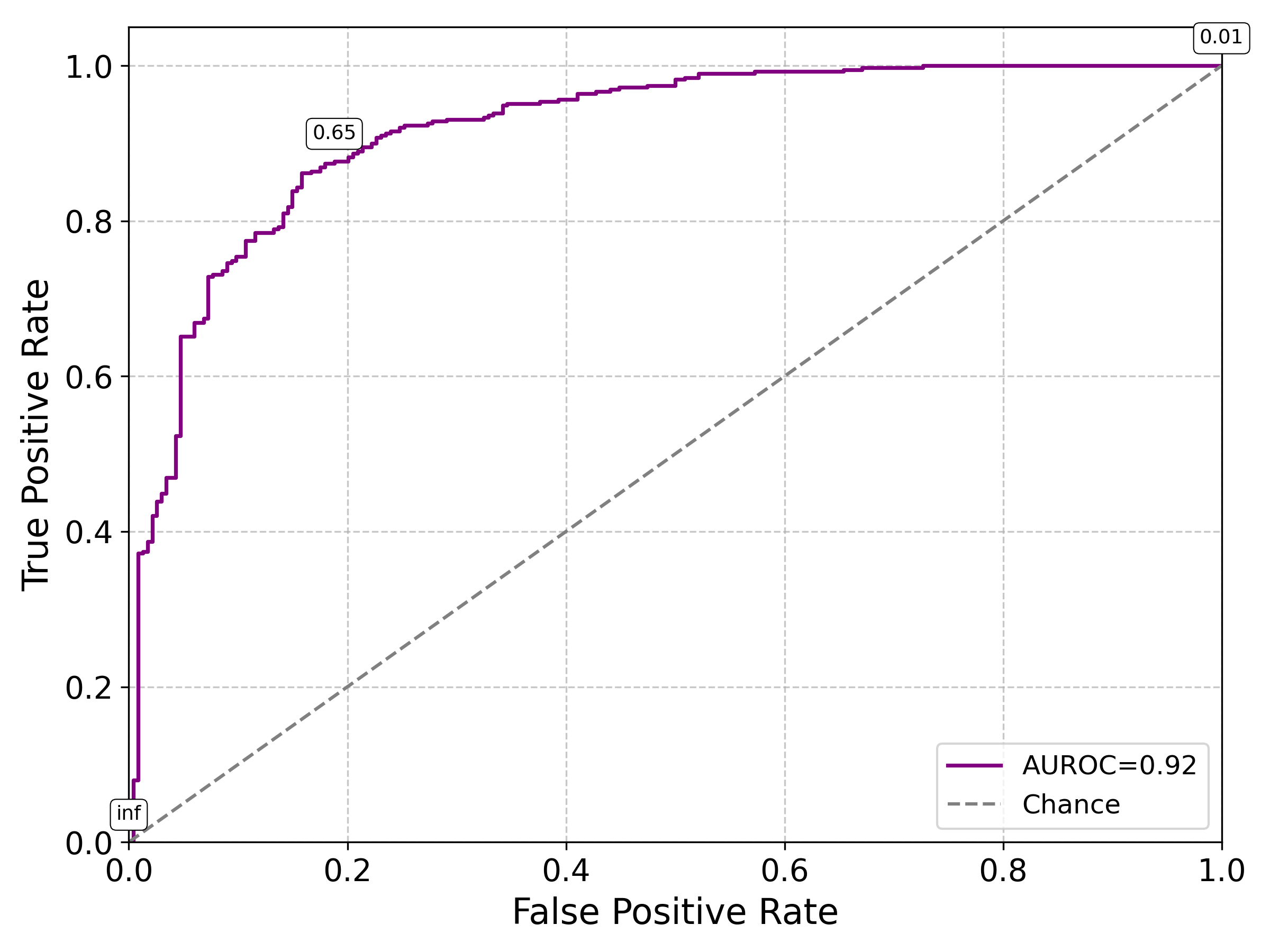}
    \hspace{0.4cm}
    \includegraphics[width=0.28\linewidth]{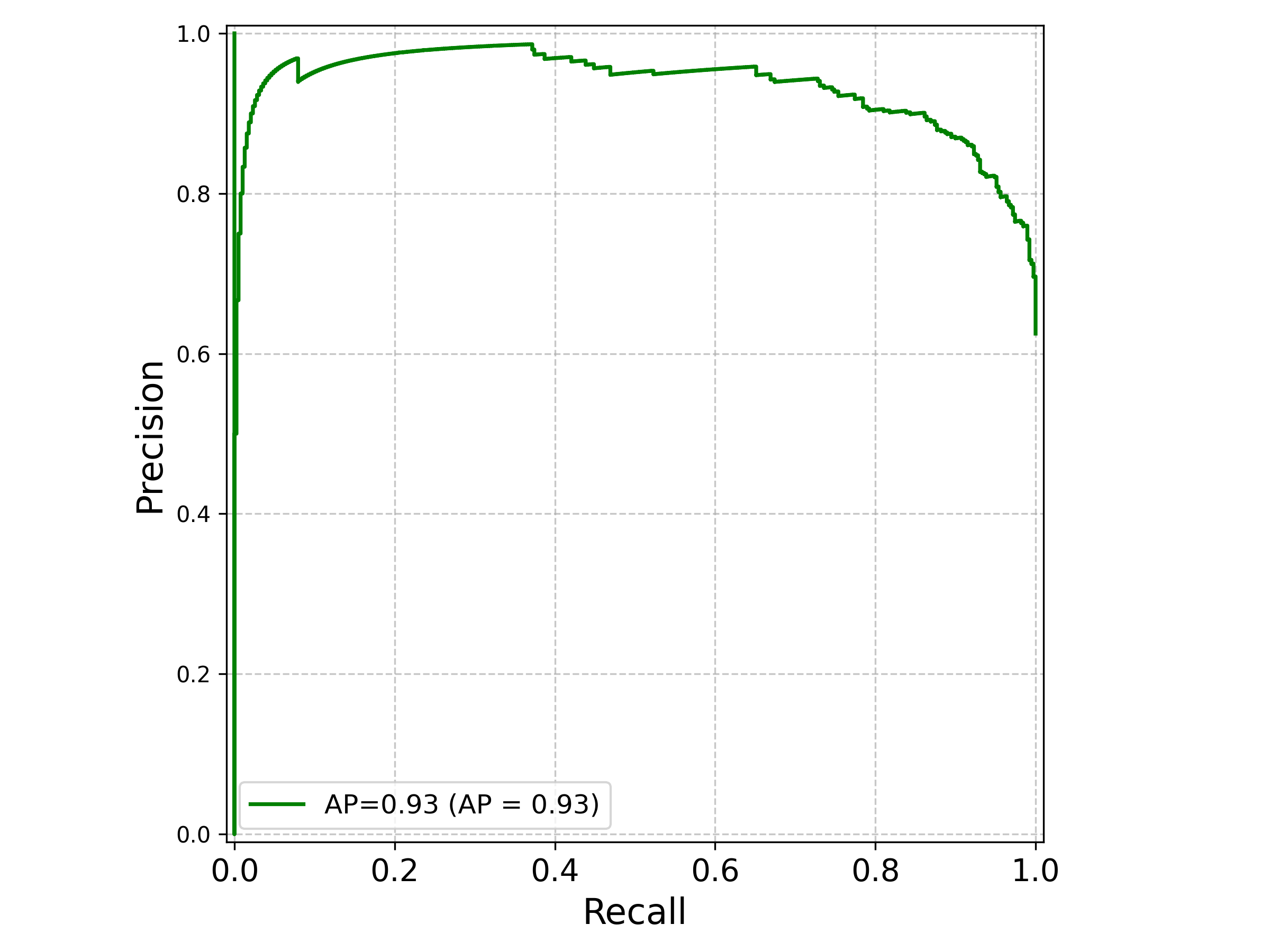}
    \hspace{0.4cm}
    \includegraphics[width=0.28\linewidth]{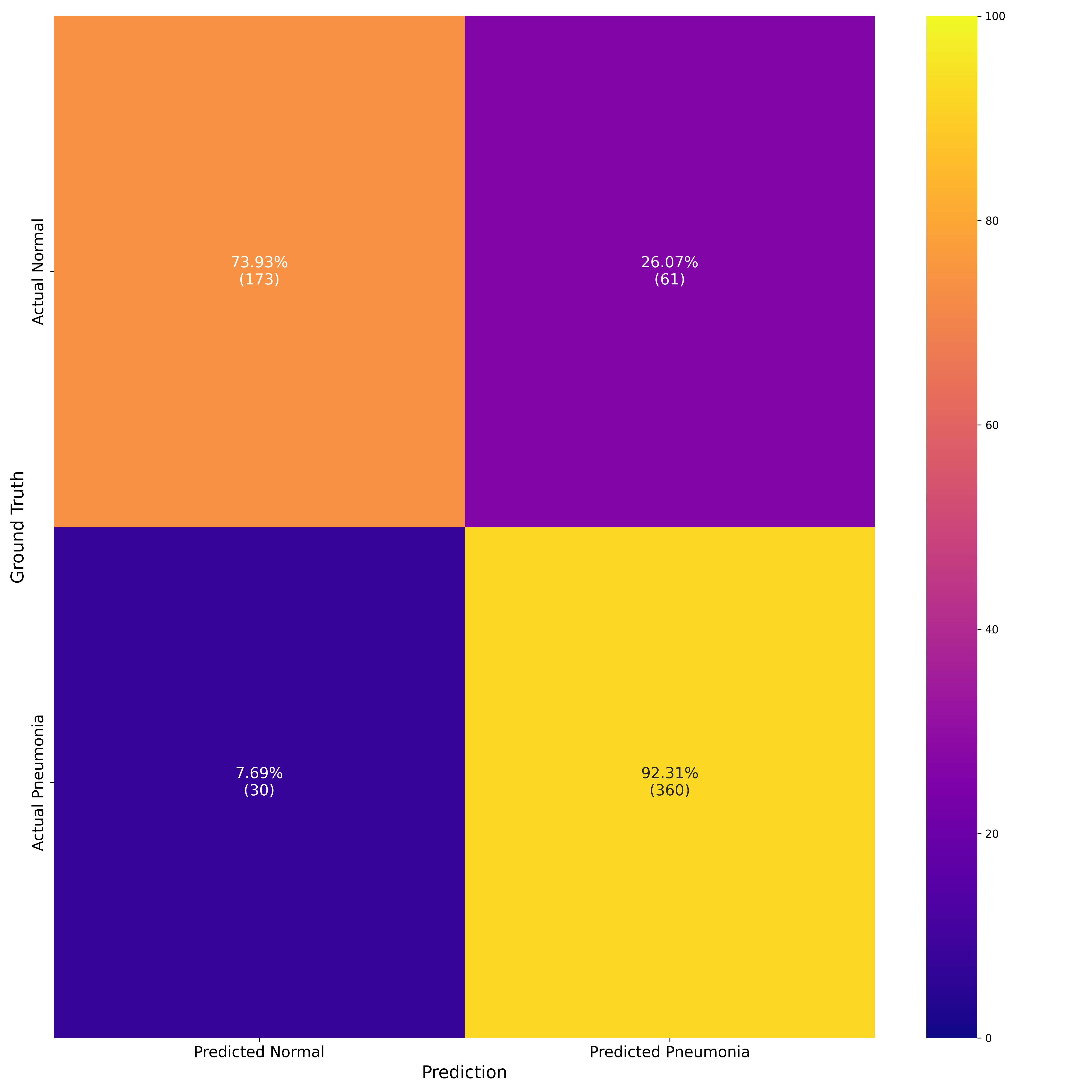}

    \caption{Performance evaluation of classical neural network on PneumoniaMNIST dataset: (a) AUROC, (b) precision-recall (PR) curve, (c) confusion matrix.}
    \label{fig:classical-nn-performance-pneumonia}
\end{figure}

To compare the performance of the proposed quantum models, a classical counterpart with an identical architecture and number of parameters is trained and evaluated over the same configuration as the quantum neural networks. For the training and validation sets, the classical neural network performed similarly to the proposed quantum models, although it showed better training stability, as well as slight improvements by attaining an average accuracy, recall, and F1 score of 91.14\%, and 91.00\% precision for the validation set. Nevertheless, once evaluated on the test set, the classification performance is comparable to the proposed quantum neural networks, attaining an average of 85.40\% for all classification metrics. This is further proven by the obtained AUROC and AUPRC shown in the plots of Figure \ref{fig:classical-nn-performance-pneumonia}, where they achieve the same values of 92\% and 93\%, respectively. Analyzing the predictions on more detail by looking at the plotted confusion matrix, the classical model achieved a higher number of TN predictions, while obtaining slightly less TP predictions.

\begin{figure}
    \centering
    \includegraphics[width=0.28\linewidth]{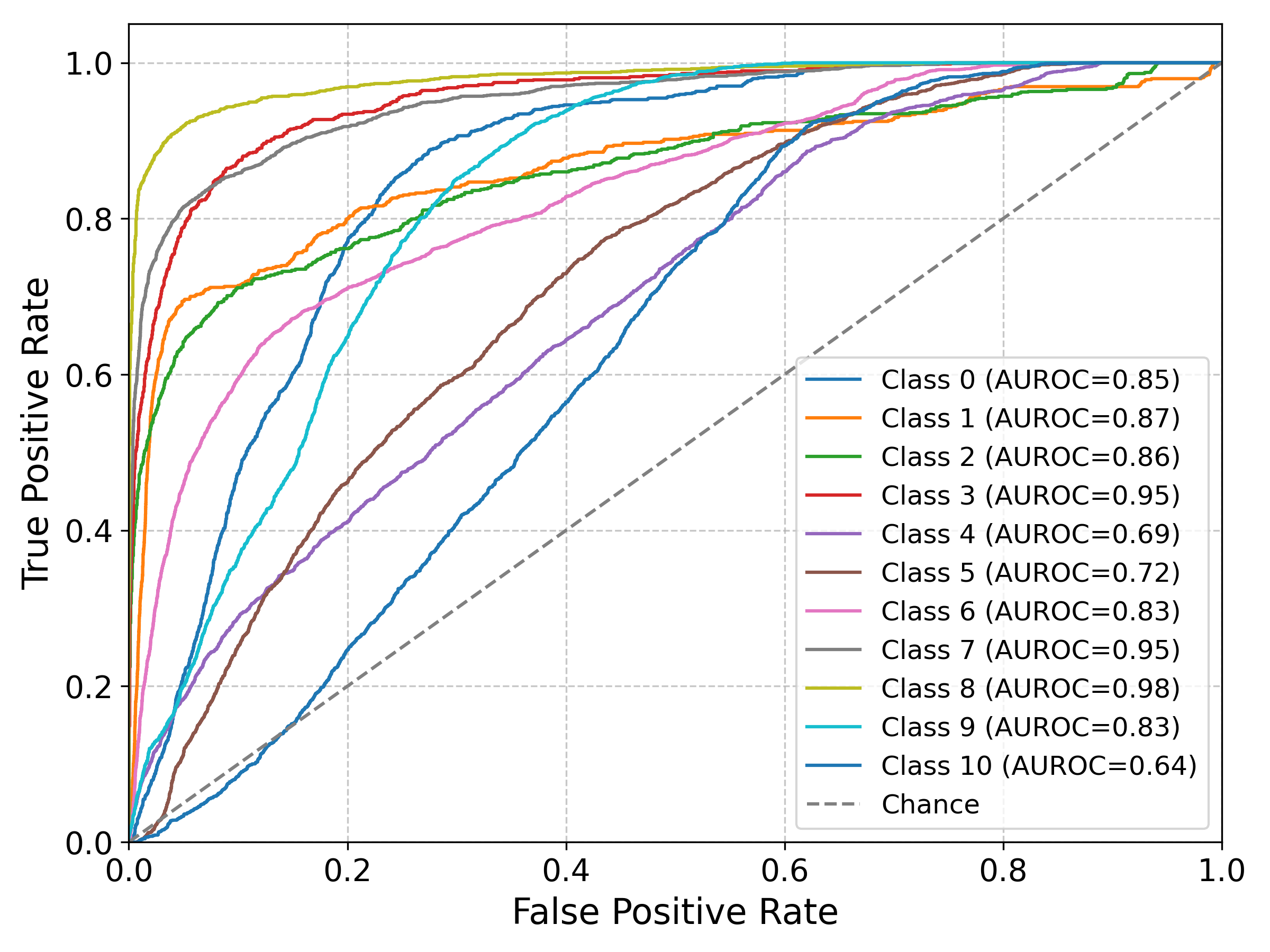}
    \hspace{0.4cm}
    \includegraphics[width=0.28\linewidth]{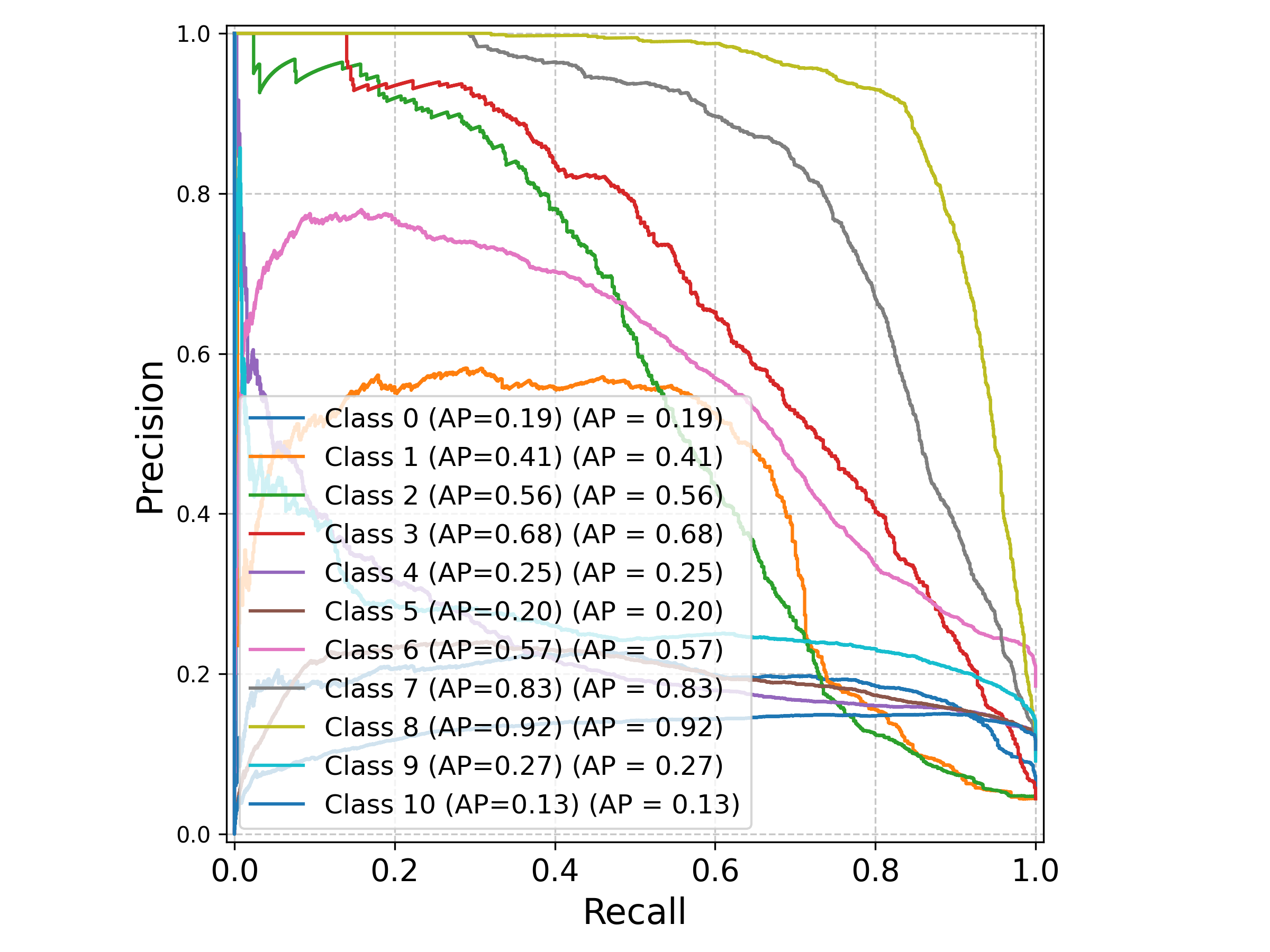}
    \hspace{0.4cm}
    \includegraphics[width=0.28\linewidth]{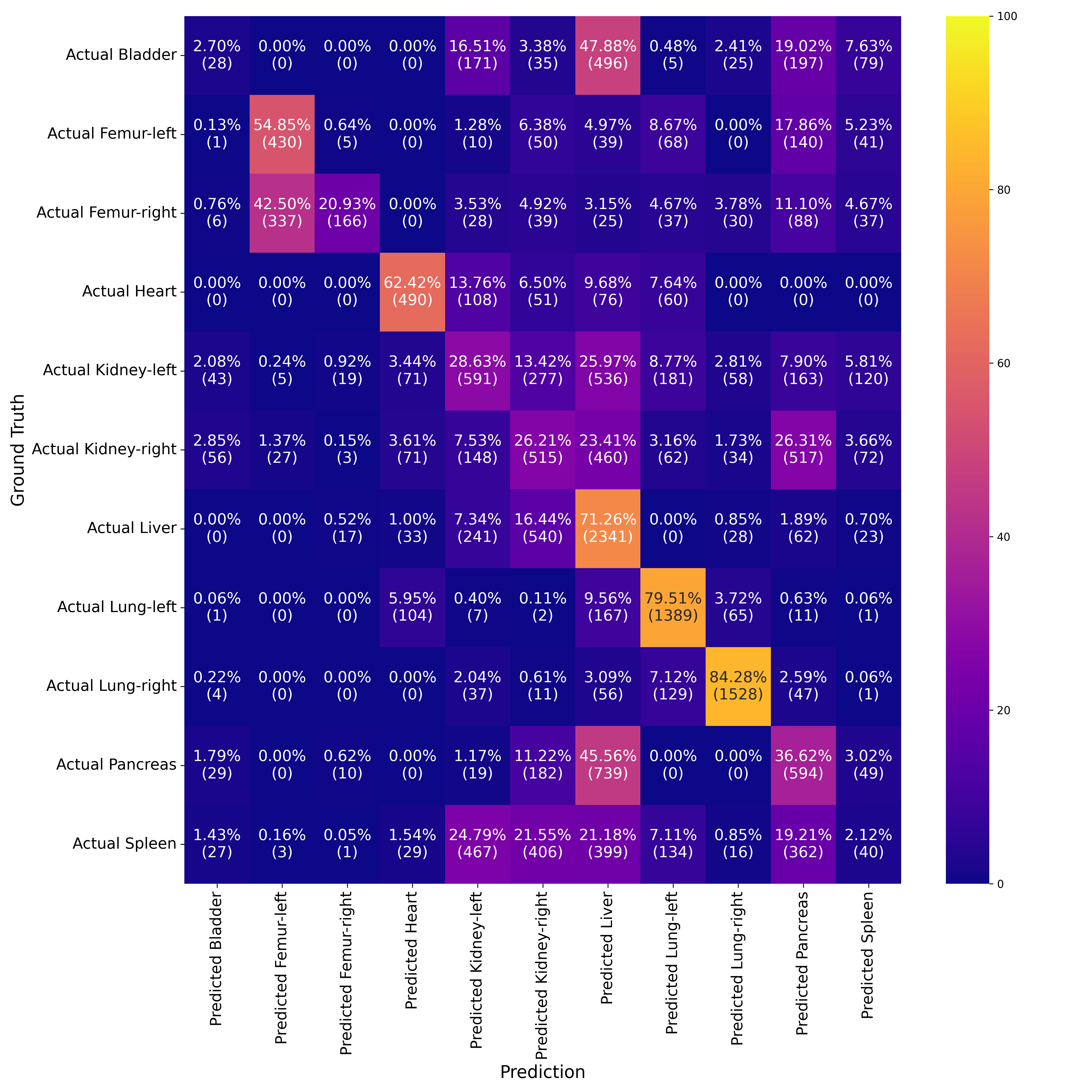}
    
    \caption{Performance evaluation of Continuous Variable Quantum Neural Network on OrganAMNIST: (a) AUROC, (b) precision-recall (PR) curve, (c) confusion matrix.}
    \label{fig:cv-qnn-performance-organa}
\end{figure}

\textbf{Experiment 2: Classification performance on OrganAMNIST dataset}\\
To evaluate the proposed CV quantum model generalization capabilities, it is trained and evaluated on the OrganAMNIST dataset comprised of 11 organ classes. Due to the increase in classes and higher degree of complexity of the dataset, the results are lower compared to the performance shown on the PneumoniaMNIST dataset. On the training set, the proposed CV quantum neural network achieves an average accuracy and of 54.20\%, and 51.36\% for precision and F1 score on the training set. Performance on the validation set is similar to the training set, with the exception of the F1 score, which increased to 53.60\%. Furthermore, evaluation on the test set showcased the model's lack of generalization for such a complex task, as it attained a test accuracy, recall and F1 score of 45.63\%, this is mainly attributed to the reduced number of parameters and data dimensionality. A more detailed look at model predictions can be seen in the plots of Figure \ref{fig:cv-qnn-performance-organa}, where the multiclass OvR AUROC, AUPRC, and confusion matrix plots are shown. Here, a class by class evaluation demonstrates that although the model attained high classification performance for the majority classes of ``Liver'', ``Lung-Left'', and ``Lung-Right'' by correctly predicting 71.26\%, 79.51\%, and 84.28\% of the class samples respectively, it still needs adjustments for the remaining organ classes. Additional class bias is further illustrated on the AUROC and AUPRC curves, where the proposed CV quantum model obtained low performance for classes 4, 5, and 10, particularly on the precision-recall curves.

\begin{figure}[ht]
    \centering
    \includegraphics[width=0.28\linewidth]{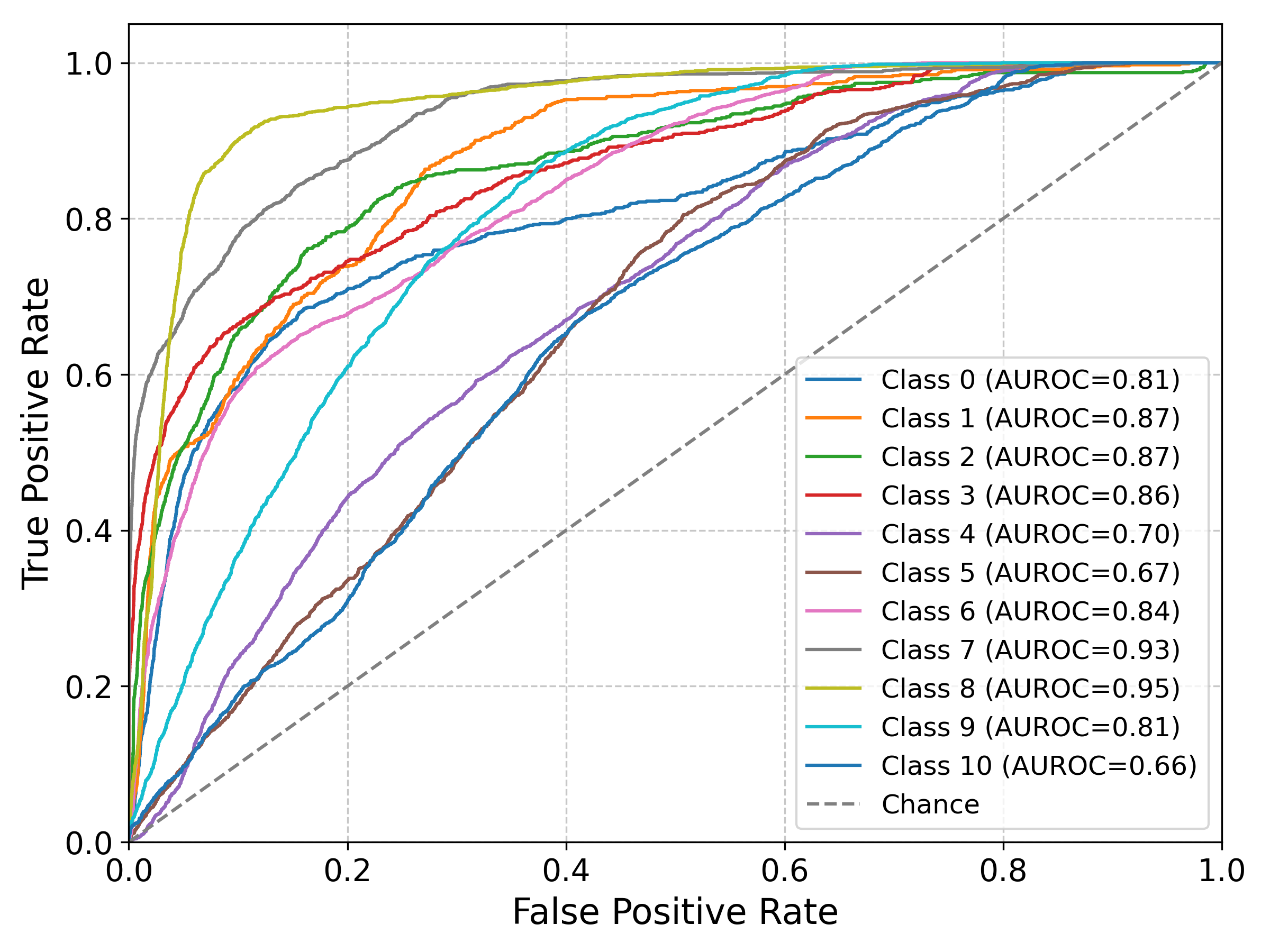}
    \hspace{0.4cm}
    \includegraphics[width=0.28\linewidth]{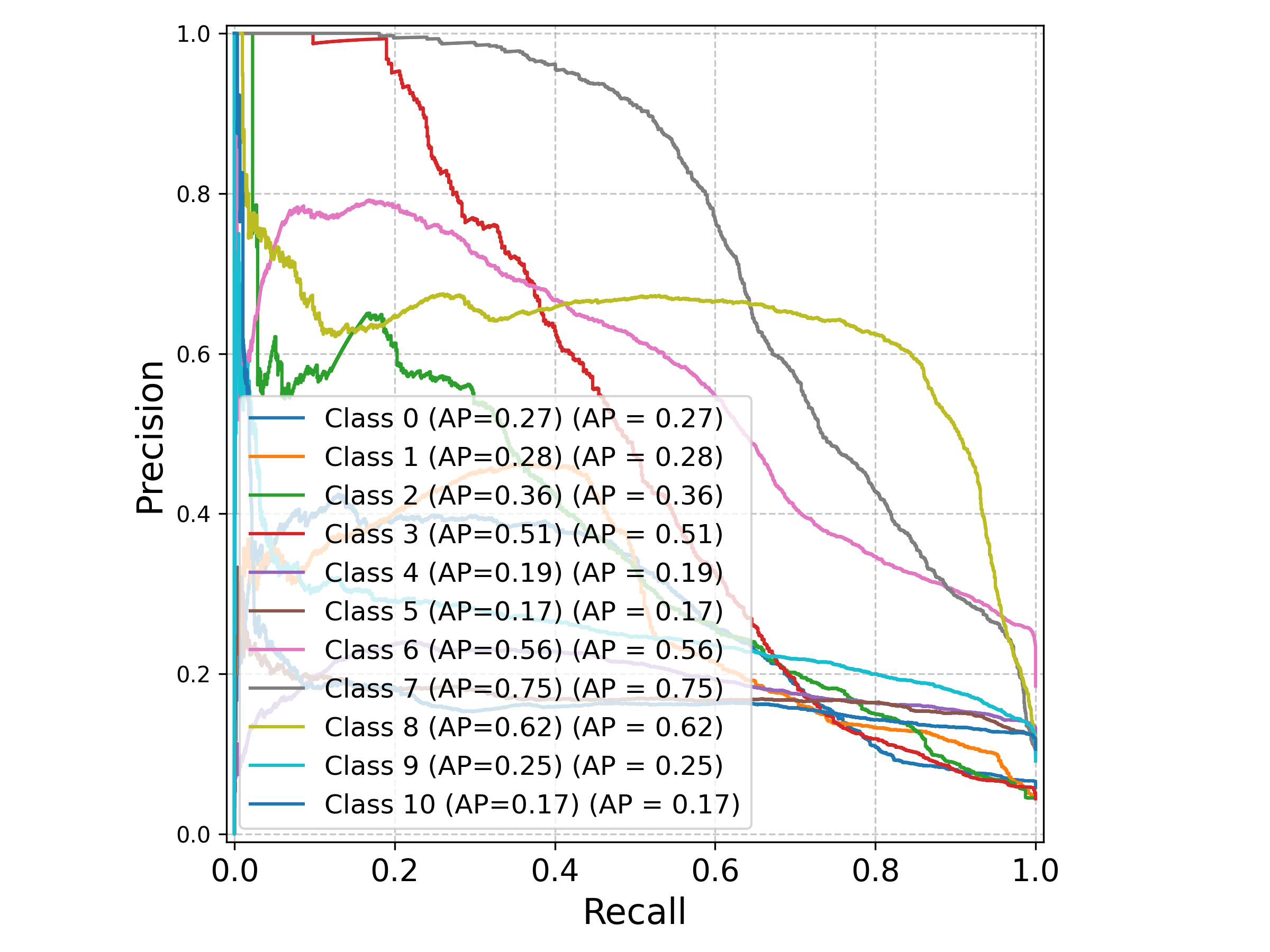}
    \hspace{0.4cm}
    \includegraphics[width=0.28\linewidth]{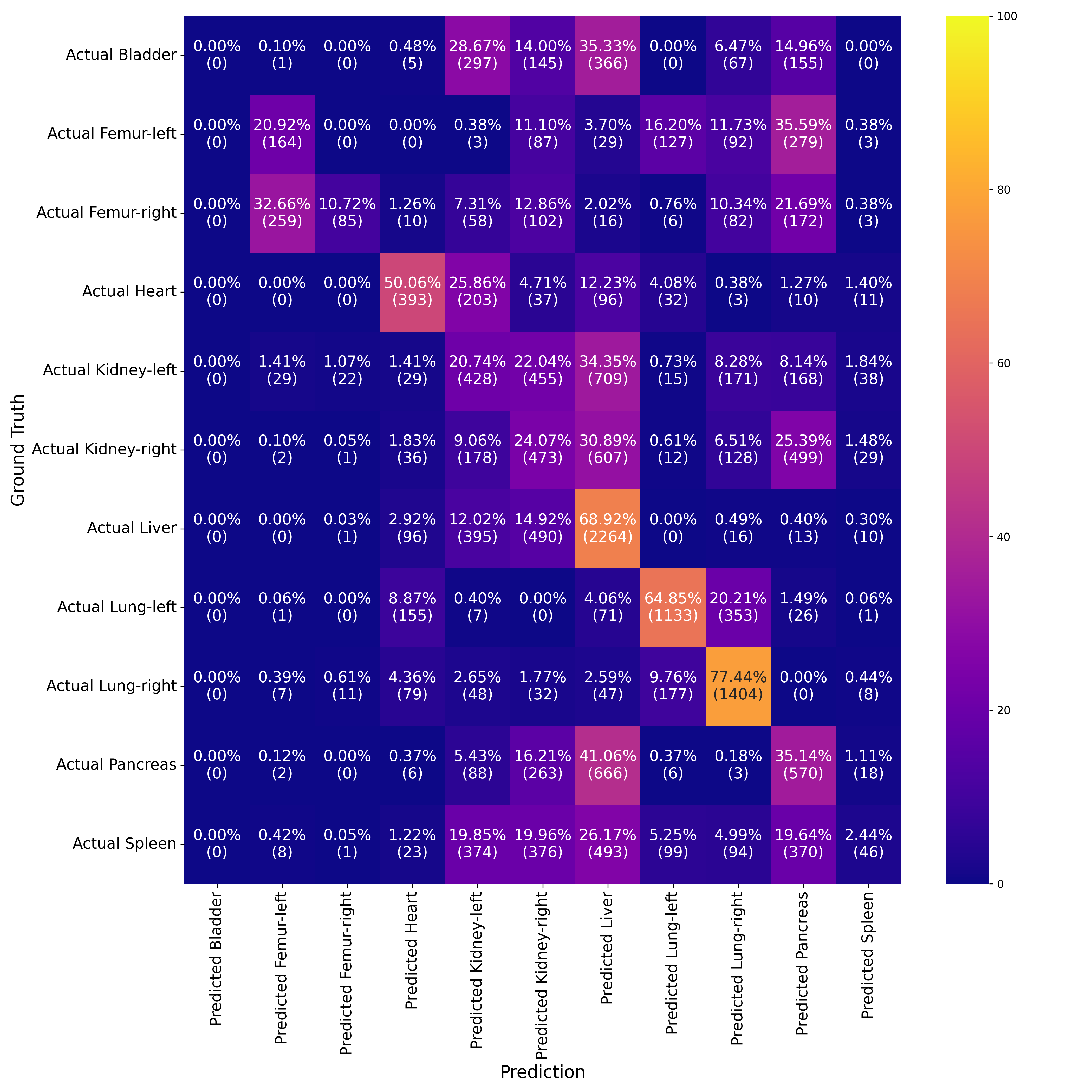}
    
    \caption{Performance evaluation of discrete variable quantum neural network on OrganAMNIST: (a) AUROc, (b) precision-recall (PR) curve, (c) confusion matrix.}
    \label{fig:dv-qnn-performance-organa}
\end{figure}

To compare the performance on multiclass classification with other methods, training and validation over the OrganAMNIST dataset for the proposed discrete variable quantum neural network is conducted. Similar to what was observed on the CV model, the increase in dataset difficulty hinders its ability to extract meaningful features and separate the 11 organ classes, resulting on a reduced average accuracy, precision, recall, and F1 score of 48.50\% for the training and validation sets. This dent in performance can also be observed on the test set results, where the DV quantum model only attained an accuracy, recall, and F1 score of 39.15\%, and a precision of 37.14\%. Moreover, as seen on the plots of Figure \ref{fig:dv-qnn-performance-organa}, the computed area under the ROC and PR curves are considerably lower, where the model struggled on classes 4, 5, and 10, being more apparent on the AUPRC. Finally, the confusion matrix plot from Figure \ref{fig:dv-qnn-performance-organa} shows that the model performed better on the majority classes ``Liver'', ``Lung-left'', and ``Lung-right'', correctly predicting 68.92\%, 64.85\%, and 77.44\%, of every class sample respectively.

\begin{figure}
    \centering
    \includegraphics[width=0.28\linewidth]{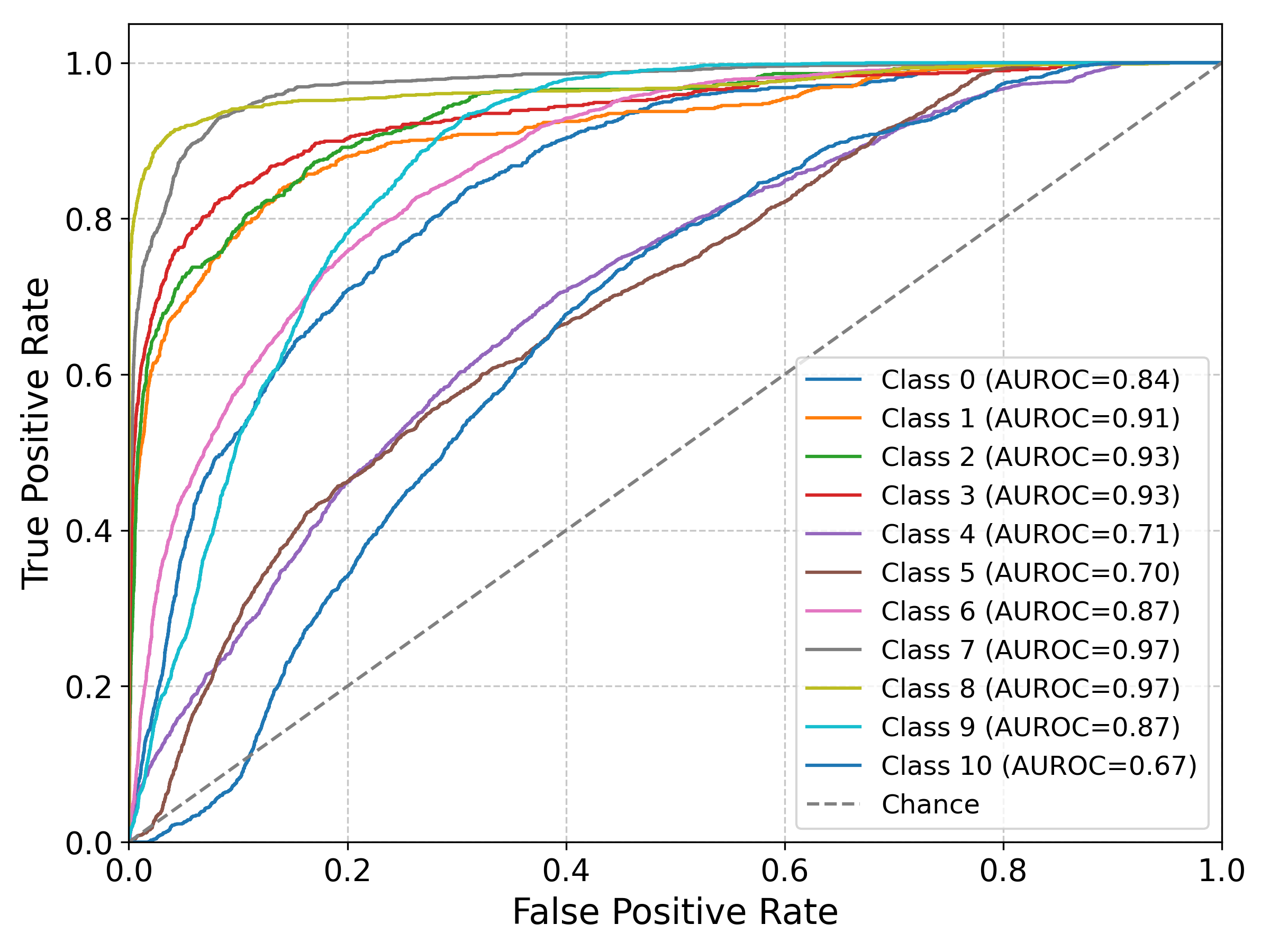}
    \hspace{0.4cm}
    \includegraphics[width=0.28\linewidth]{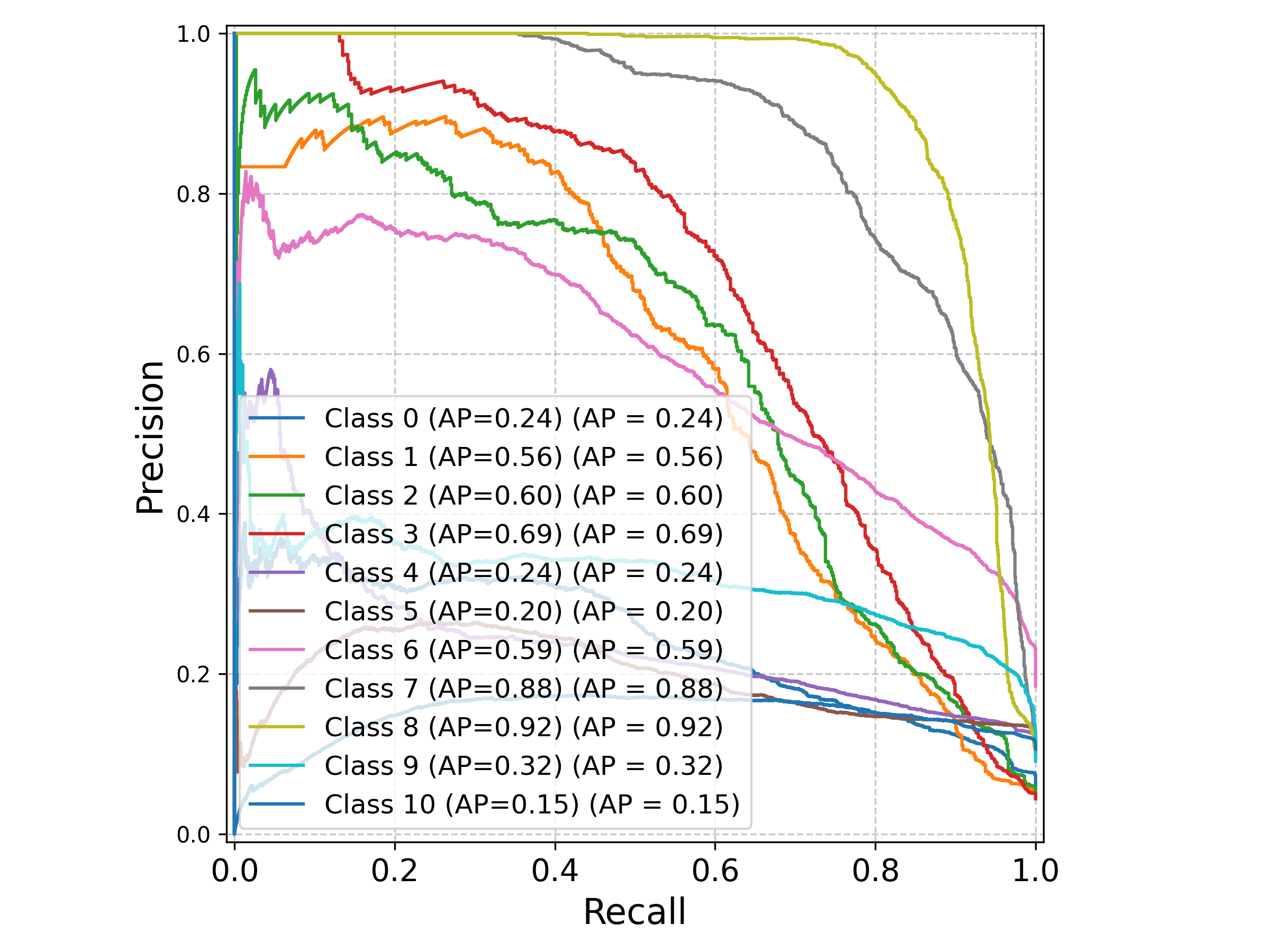}
    \hspace{0.4cm}
    \includegraphics[width=0.28\linewidth]{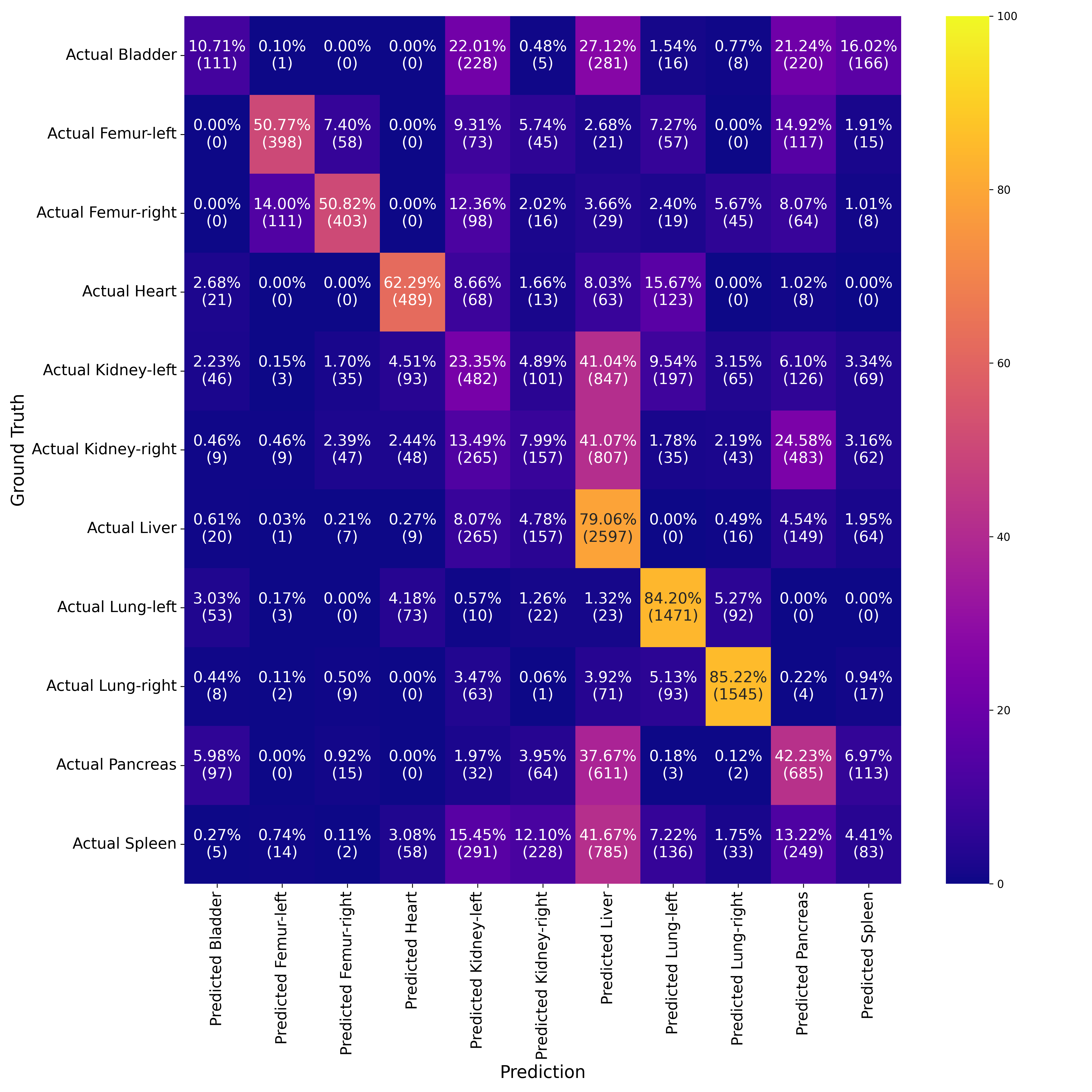}
    
    \caption{Performance evaluation of classical neural network on OrganAMNIST: (a) AUROC, (b) precision-recall (PR) curve, (c) confusion matrix.}
    \label{fig:classical-nn-performance-organa}
\end{figure}

Likewise with the PneumoniaMNIST dataset, the classical model is also used as reference to the proposed quantum models. Over the 50 epochs of threefold cross-validation, the classical model attained an average accuracy, recall, and F1 score of 55.25\%, and an average precision of 49.58\% for the training set, while achieving 54.92\% average accuracy, recall, and F1 score, and 49.85\% average precision for the validation set, slightly overperforming when compared to its quantum counterparts. For the test set evaluation, the classical model showcased slight advantage over the CV quantum model, obtaining 47.37\% accuracy, recall, and F1 score, and 43.55\% precision, demonstrating difficulties for TP values just as the proposed quantum models. More thorough analysis over various thresholds is shown in Figure \ref{fig:classical-nn-performance-organa} on the AUROC and AUPRC curves, achieving a small increase in performance compared to its quantum counterparts, particularly on the AUPRC curves for classes 4, 5, 8, and 9. Lastly, the classical model predictions showed in the confusion matrix of Figure \ref{fig:classical-nn-performance-organa} further proves its slight advantage over the quantum models in most predictions. 

\begin{figure}
    \centering
    \includegraphics[width=0.28\linewidth]{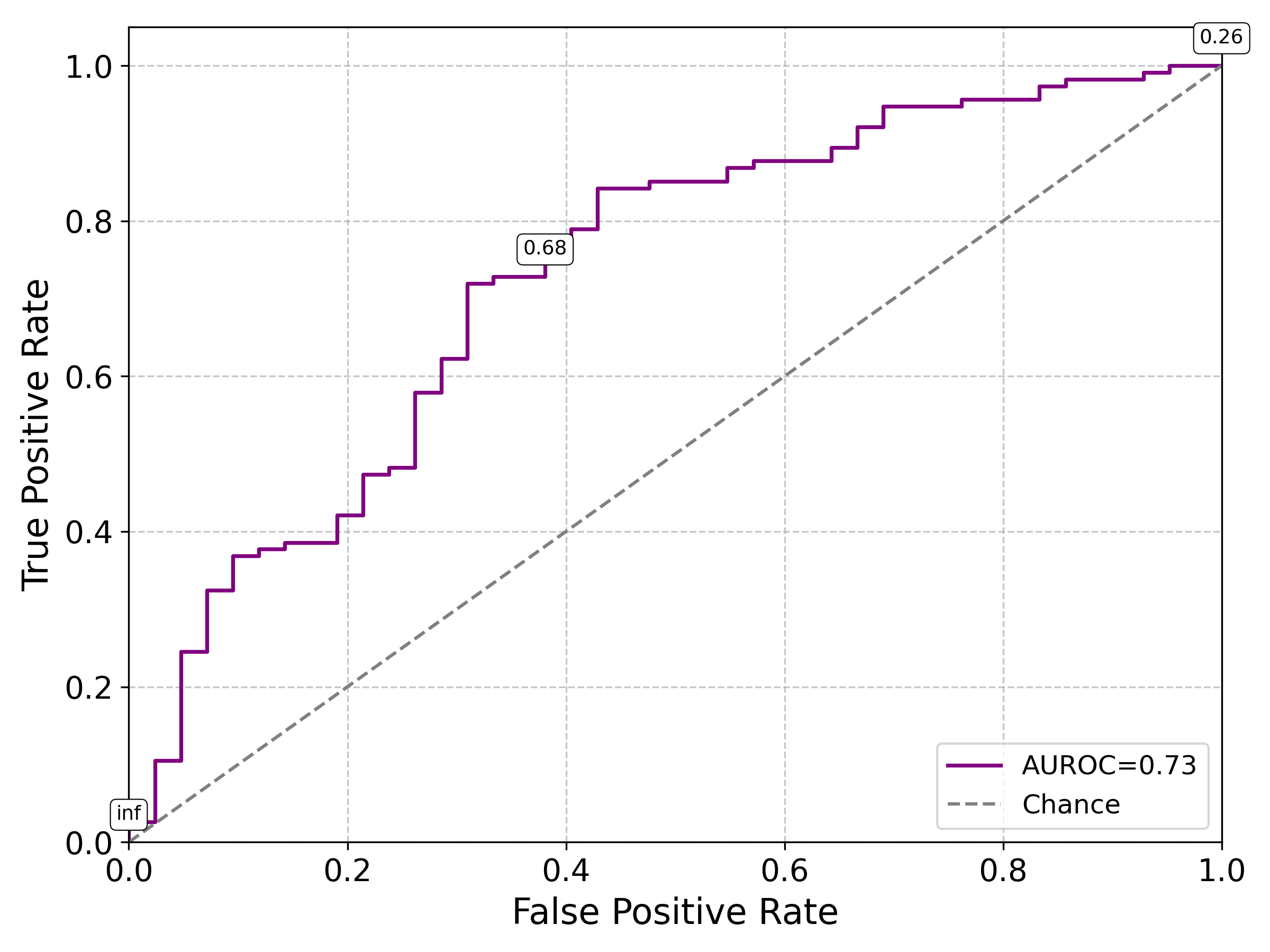}
    \hspace{0.4cm}
    \includegraphics[width=0.28\linewidth]{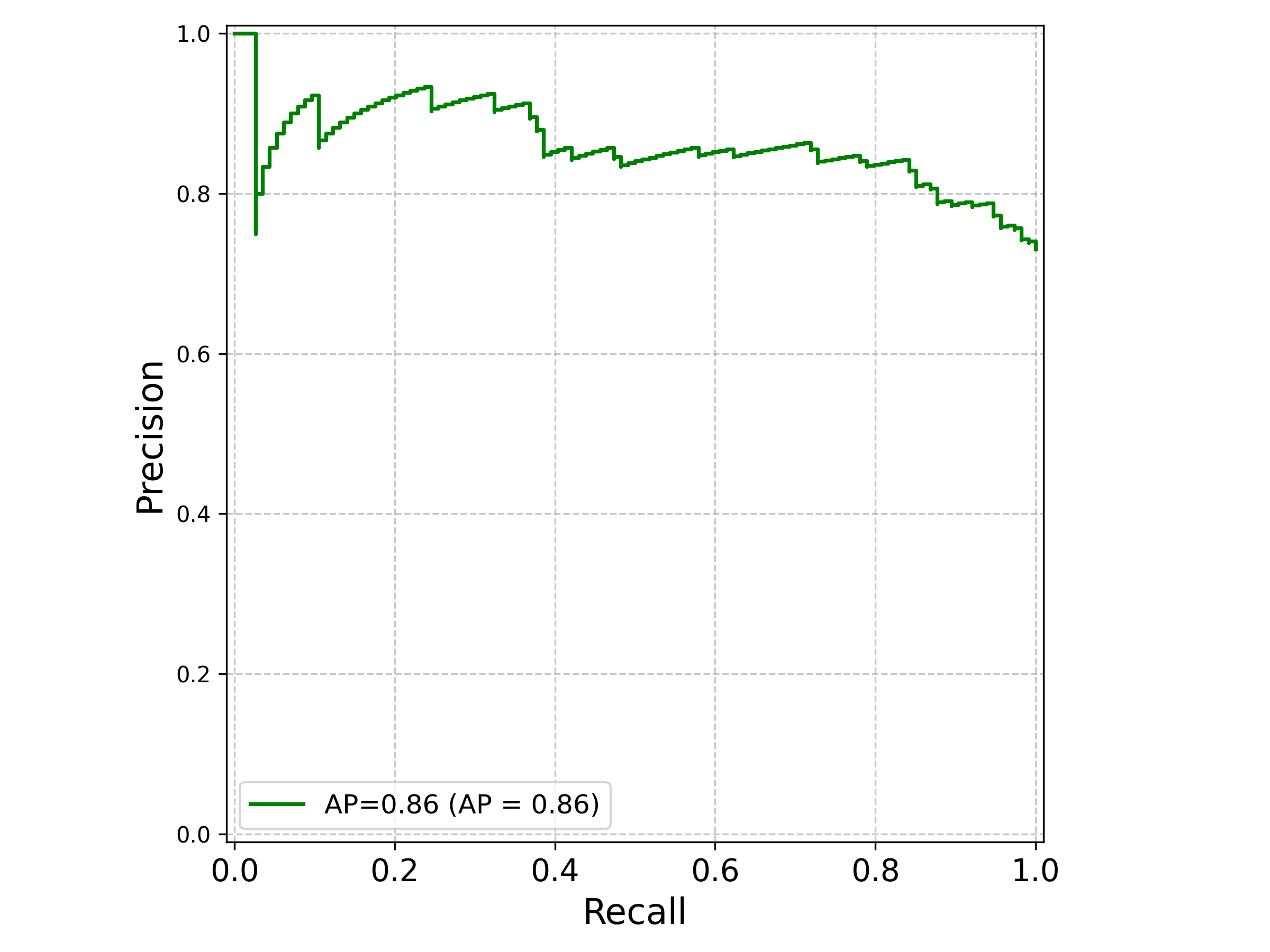}
    \hspace{0.4cm}
    \includegraphics[width=0.28\linewidth]{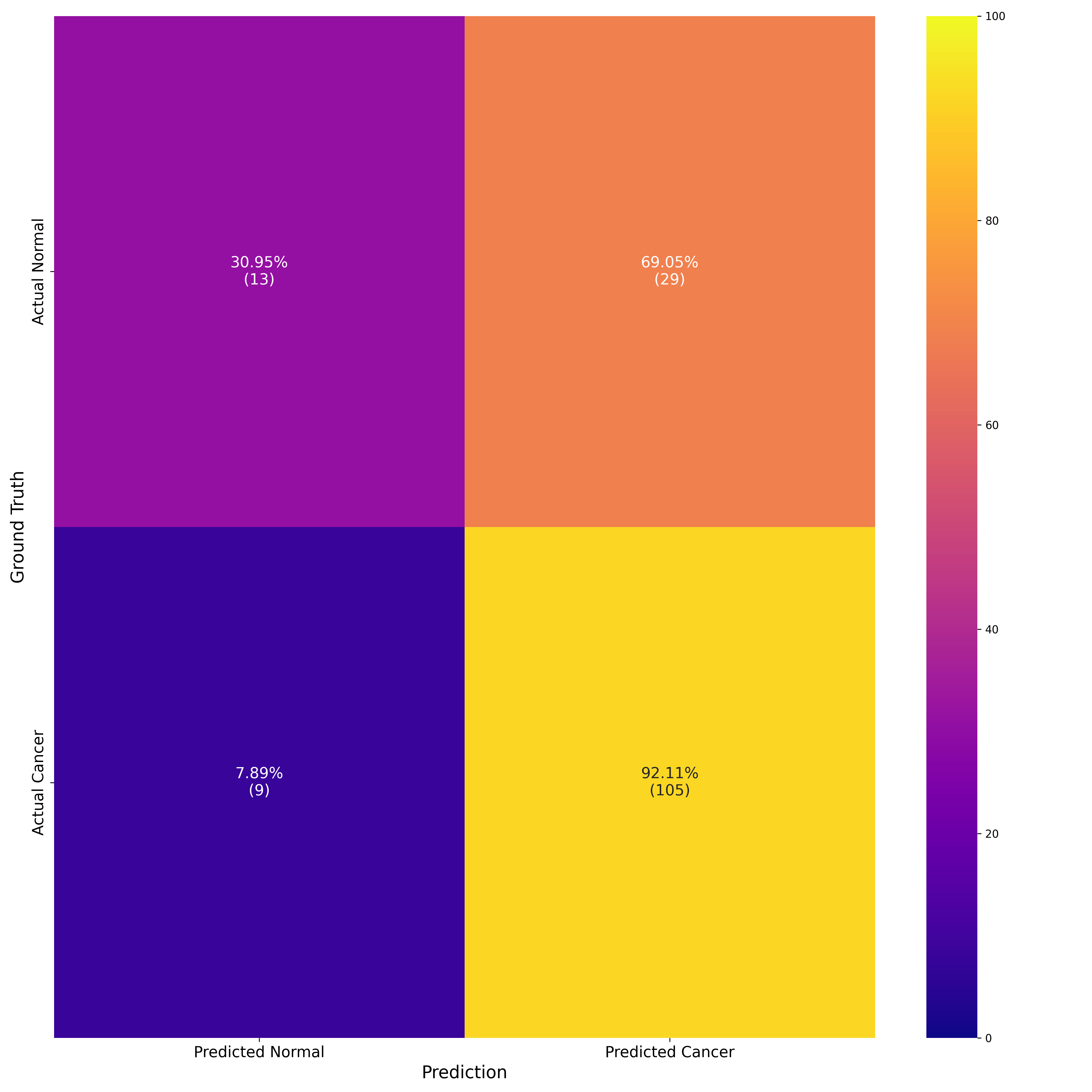}
    
    \caption{Performance evaluation of Continuous Variable Quantum Neural Network on BreastMNIST: (a) AUROC, (b) precision-recall (PR) curve, (c) confusion matrix.}
    \label{fig:cv-qnn-performance-breast}
\end{figure}

\textbf{Experiment 3: Classification performance on BreastMNIST dataset}\\
To conclude classification performance, the proposed quantum models and their classical counterpart are trained and evaluated on the BreastMNIST, a small and imbalanced binary classification dataset, where the models' focus on minority classes is assessed. The proposed continuous variable quantum neural network attained an average accuracy, recall, and F1 score of 72.34\% for the train and validation sets, and an average precision of 70.82\%, demonstrating favorable performance despite the reduced data dimensionality and number of parameters. These results are further validated by the results on the test set, where it achieves 75.64\% of accuracy, recall, and F1 score, and a precision of 73.17\%. Moreover, the proposed CV quantum model attains an area under the ROC of 73\%, and an area under the PR curve of 86\% as shown in Figure \ref{fig:cv-qnn-performance-breast}, displaying favorable generalization despite data dimensionality reduction and small a set of trainable parameters. However, the plotted confusion matrix illustrates the issue of the CV quantum model in correctly identifying the negative class, as it only correctly predicts 30.95\% of them.

\begin{figure}[ht]
    \centering
    \includegraphics[width=0.28\linewidth]{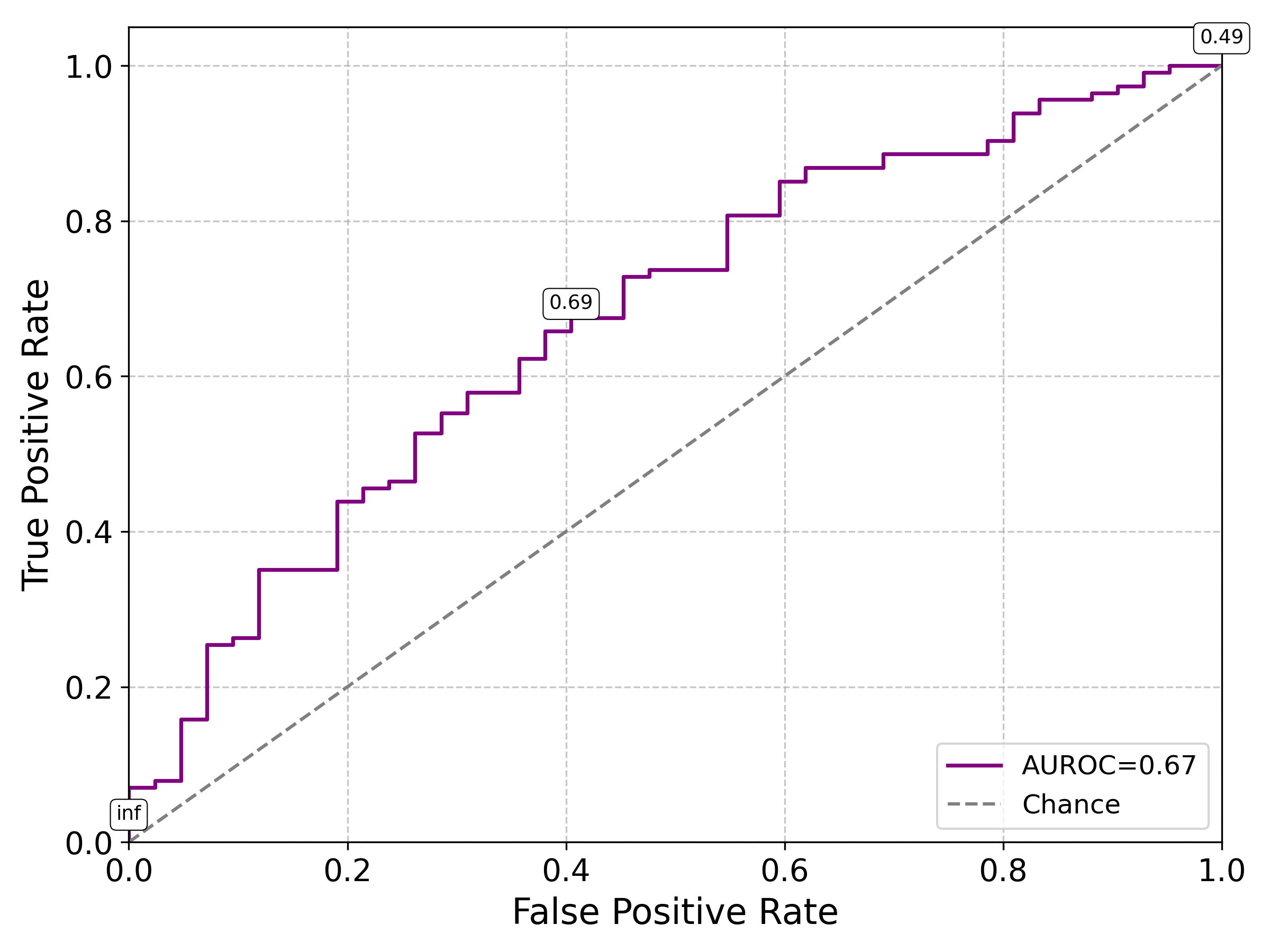}
    \hspace{0.4cm}
    \includegraphics[width=0.28\linewidth]{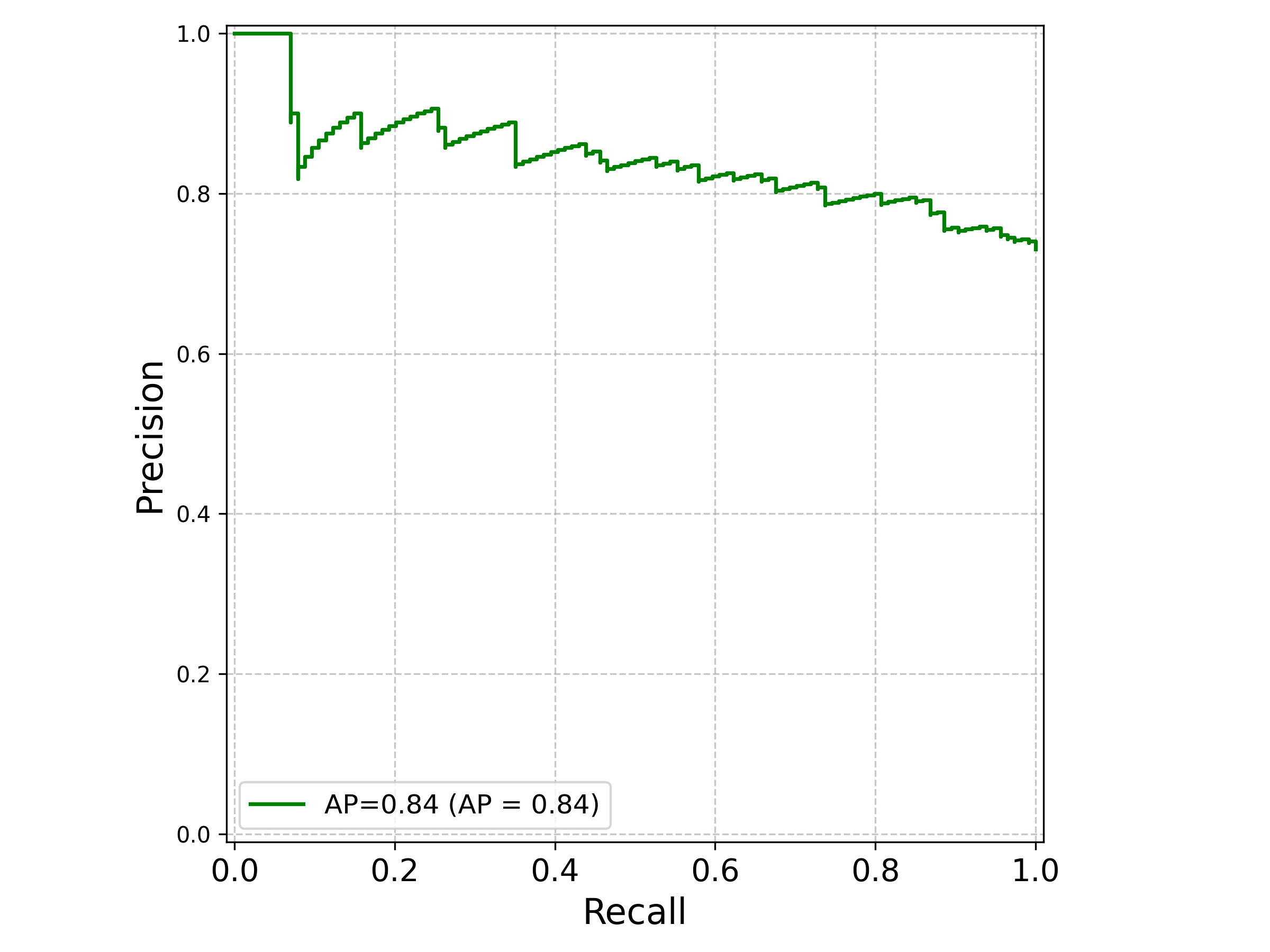}
    \hspace{0.4cm}
    \includegraphics[width=0.28\linewidth]{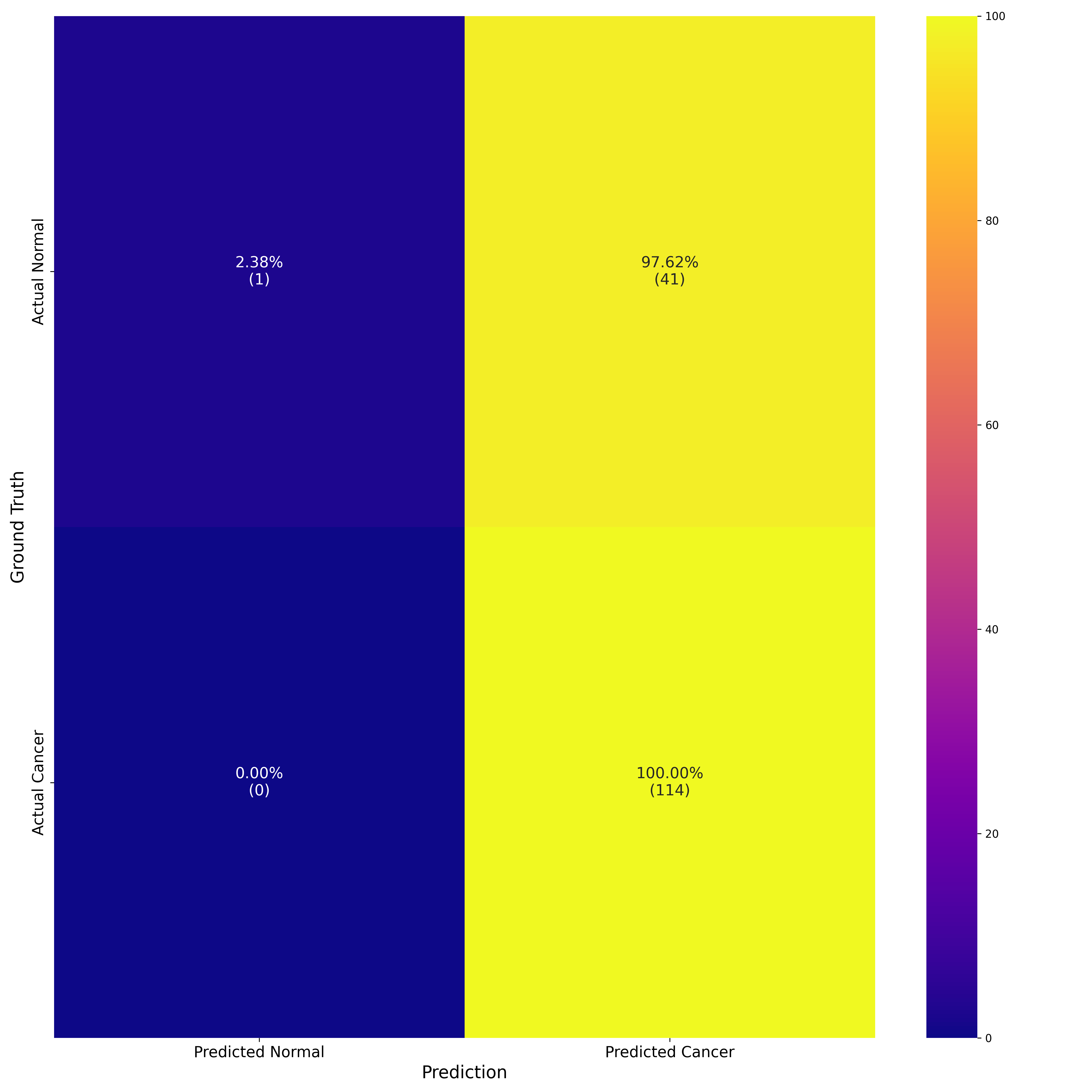}
    
    \caption{Performance evaluation of discrete variable quantum neural network on BreastMNIST: (a) AUROC, (b) precision-recall (PR) curve, and (c) confusion matrix.}
    \label{fig:dv-qnn-performance-breast}
\end{figure}

The proposed discrete variable quantum neural network is also trained and evaluated on the BreastMNIST dataset, and attained a favorable but less stable performance on the training and validation sets compared to its CV counterpart. The proposed DV quantum model achieved an average accuracy and recall of 73.83\% for the training set, and a slightly lower 72.17\% for the validation set. However, instabilities are more apparent for precision an F1 score, where a bigger discrepancy between training and validation set results is apparent. Here, the DV quantum model reached an average precision of 72.16\%, and an F1 score of 73.46\% for the training set, but for validation it achieved an average of 68.32\% and 65.67\%, respectively. Nevertheless, during test set evaluation the DV quantum model attained an accuracy, recall, and F1 score of 73.72\%, a similar although slightly lower when compared to its CV counterpart, but an increased precision of 80.67\% compared to the proposed CV quantum model. These results are apparent when looking at the plots of Figure \ref{fig:dv-qnn-performance-breast}, where the model achieved high classification performance for positive samples as shown in the confusion matrix where it correctly identified 100\% of the breast cancer samples, but struggled with the negative samples. The computed area under the ROC and PR curves also displays this behavior, as it only reaches 67\% and 84\% when evaluated over various decision thresholds.

\begin{figure}[ht]
    \centering
    \includegraphics[width=0.28\linewidth]{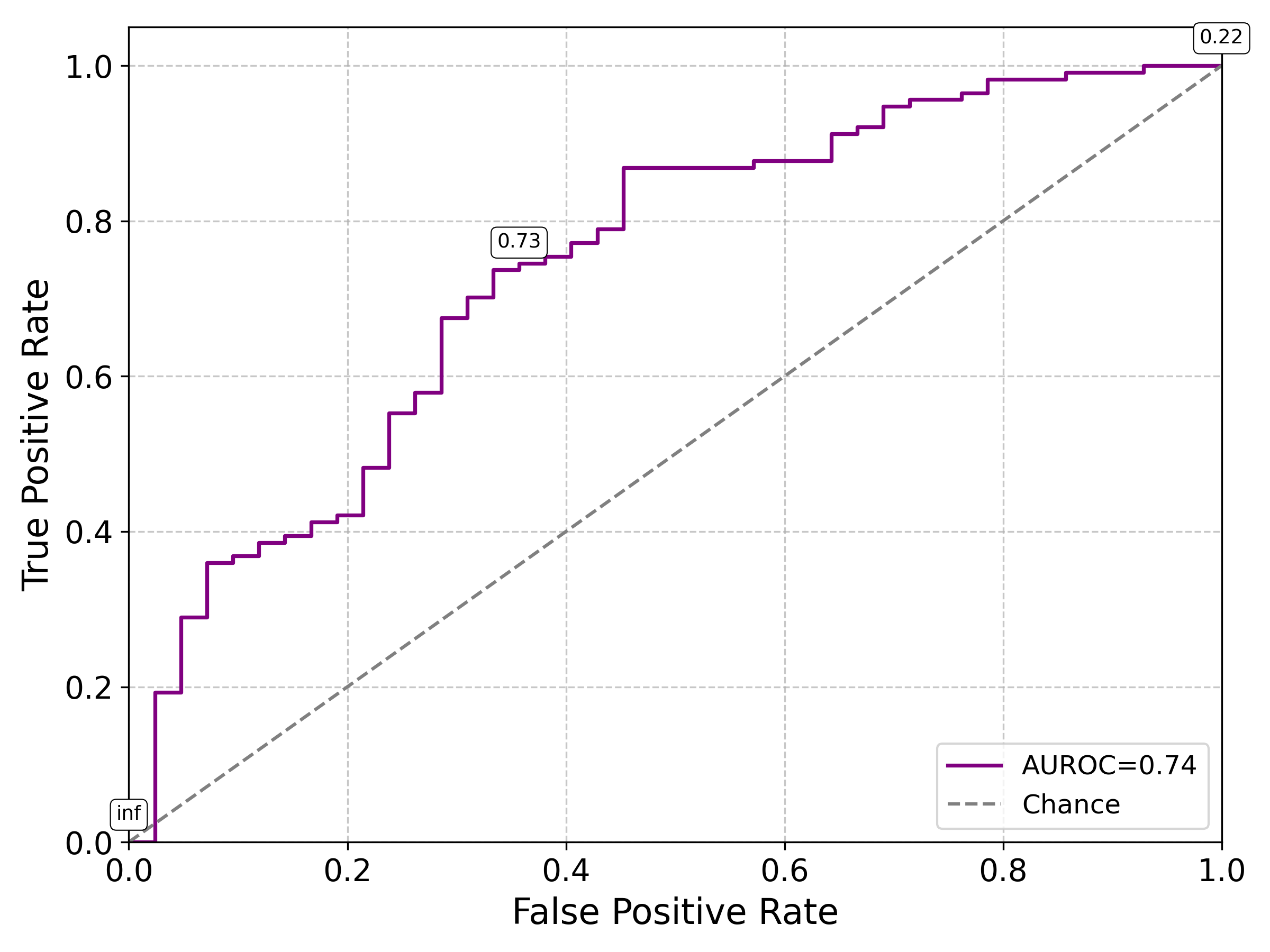}
    \hspace{0.4cm}
    \includegraphics[width=0.28\linewidth]{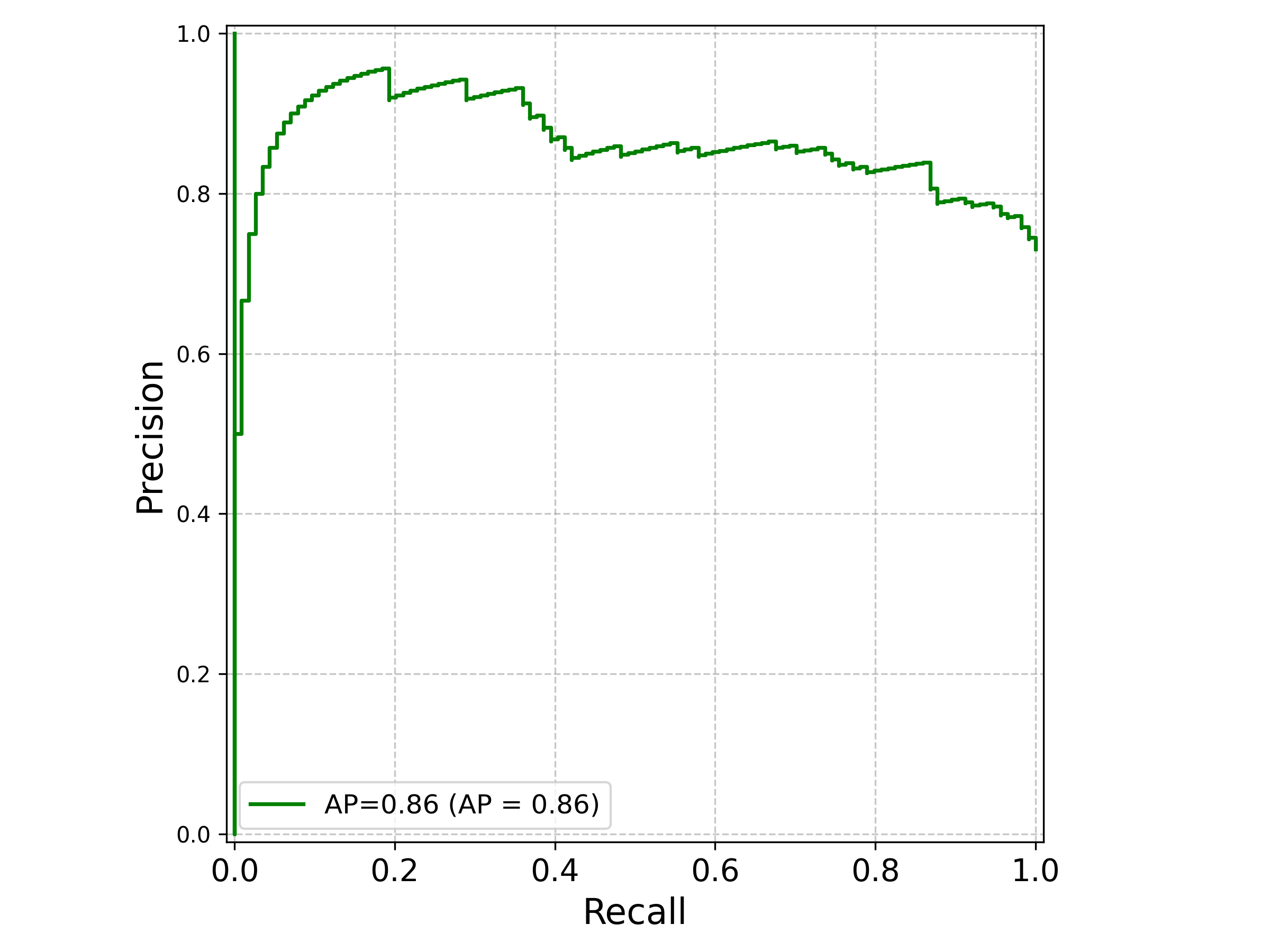}
    \hspace{0.4cm}
    \includegraphics[width=0.28\linewidth]{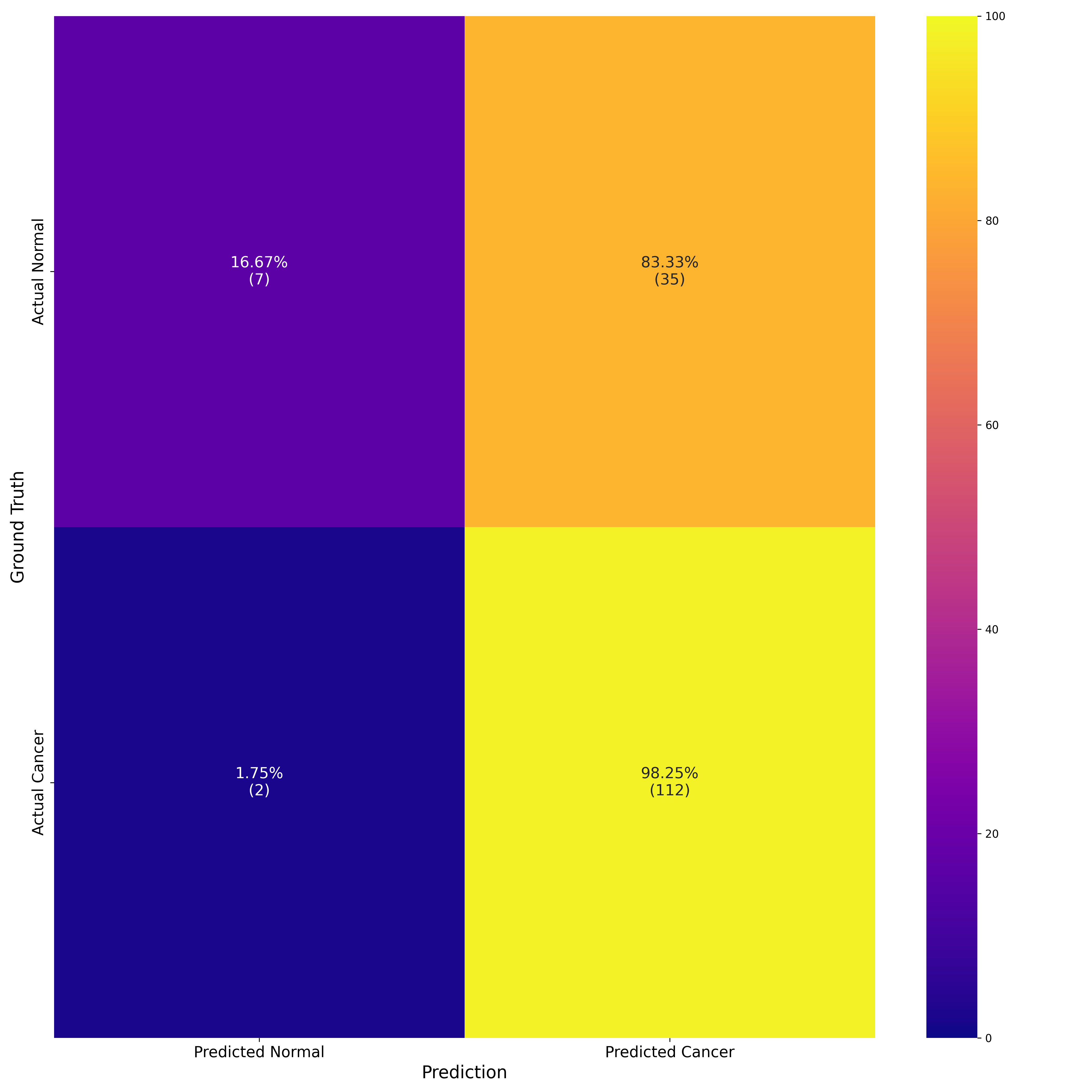}
    
    \caption{Performance evaluation of classical neural network on BreastMNIST: (a) AUROC, (b) precision-recall (PR) curve, and (c) confusion matrix.}
    \label{fig:classical-nn-performance-breast}
\end{figure}

The classical model is also trained and evaluated on the BreastMNIST dataset for benchmarking comparison with the proposed models. Similarly to what was obtained on the previous datasets, the classical model showcased a slight advantage over the proposed quantum models, attaining an average accuracy and recall of 75.82\%, a precision of 74.42\%, and an F1 score of 70.25\% for the training set. Likewise, the classical model attains an average accuracy, recall, and F1 score of 74.73\%, and a precision of 71.73\% for the validation set. Furthermore, test set evaluation results showcased the classical model generalization capabilities by achieving 76.28\% on all classification metrics. Nonetheless, its capability to distinguish negative samples and in consequence minority classes is lower than the proposed CV quantum model, and although it achieved an AUROC and AUPRC of 74\% and 86\%, respectively, the number of correctly identified negative samples is half of what the CV quantum model attained, as shown in the plots of Figure \ref{fig:classical-nn-performance-breast}. 

\begin{figure}[ht]
    \centering
    \includegraphics[width=0.28\linewidth]{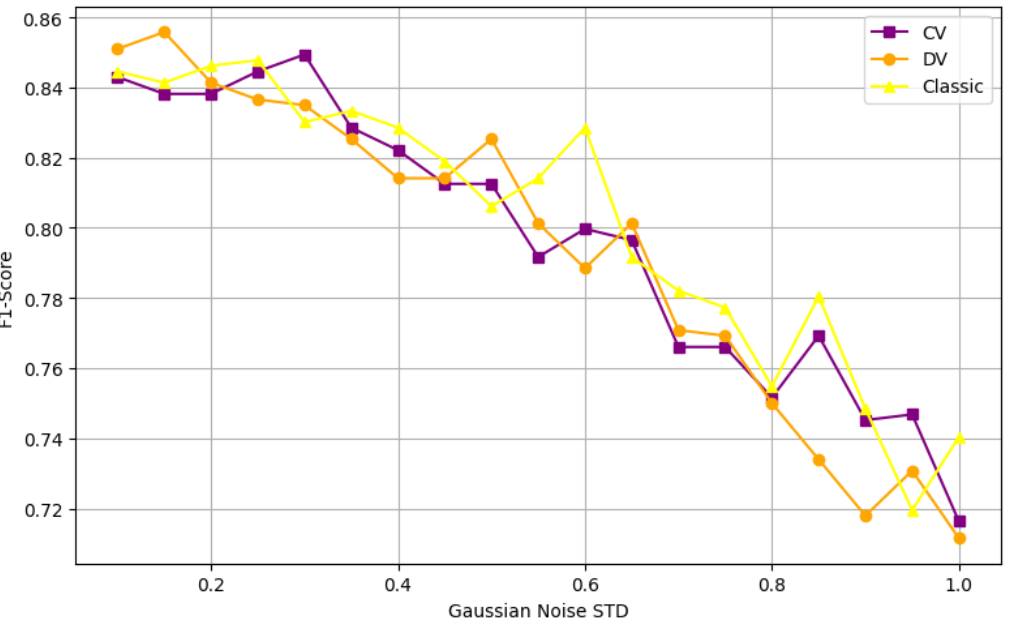}
    \hspace{0.4cm}
    \includegraphics[width=0.28\linewidth]{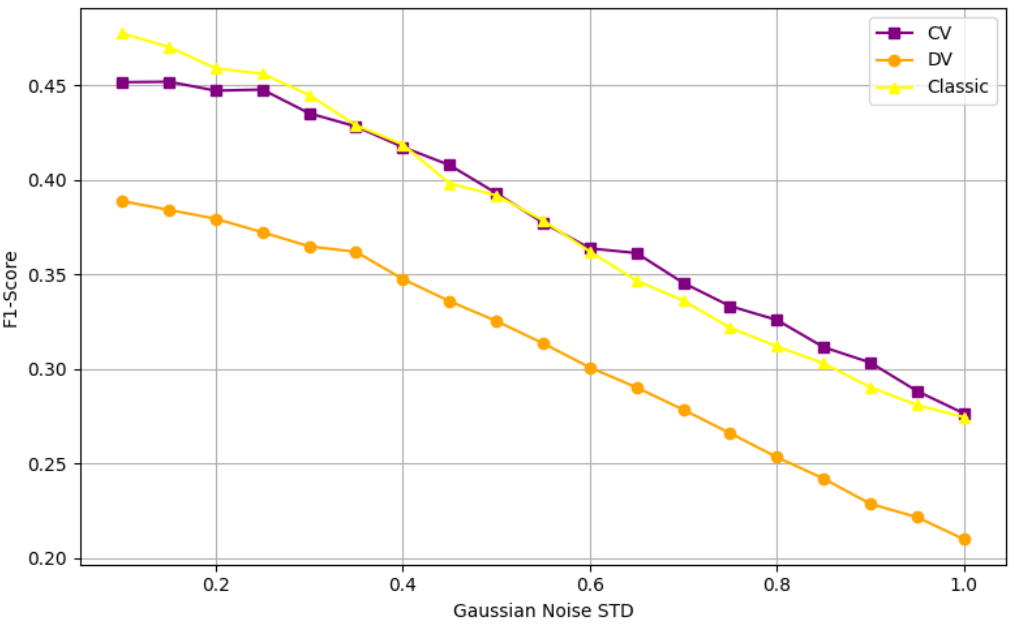}
    \hspace{0.4cm}
    \includegraphics[width=0.28\linewidth]{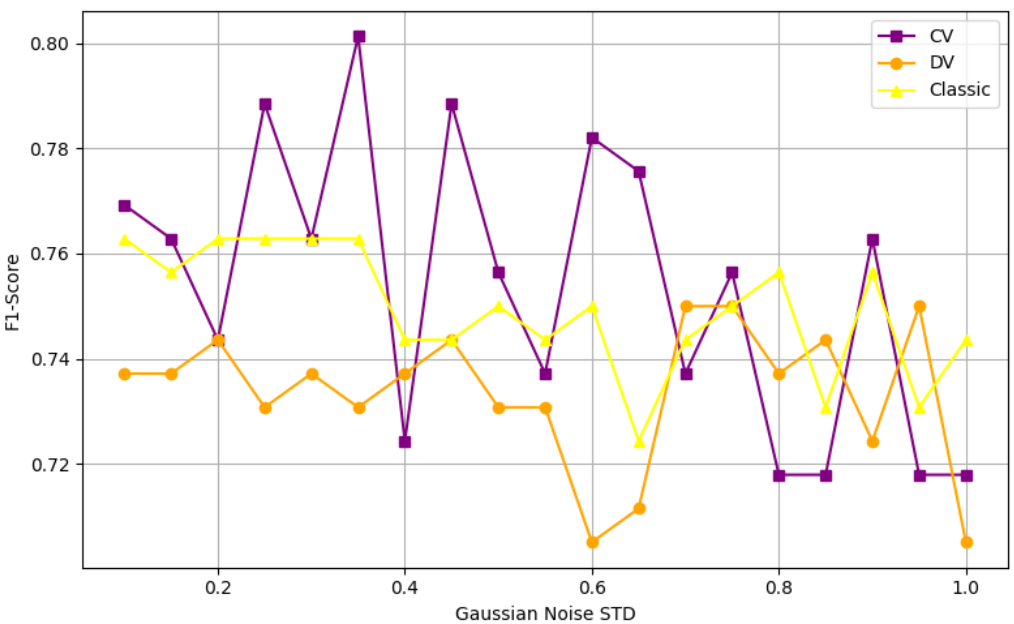}
    
    \caption{Noise robustness comparison between models for every considered dataset of MedMNIST database.}
    \label{fig:noise-robust-comparison}
\end{figure}

\textbf{Experiment 4: Noise robustness model comparison}\\
To assess robustness of the proposed quantum neural networks, we test the trained models on test set classification over different levels of random Gaussian noise. Random Gaussian noise ranging from [0.1, 1.0] with increments of 0.05 is injected to the test set images, following this, evaluation is conducted for the proposed quantum models, as well as the classical model. The set of plots shown in Figure \ref{fig:noise-robust-comparison} displays the behavior observed for every model. The plot on (a) displays noise robustness over the PneumoniaMNIST dataset, where a similar F1 score for all models is shown, however, the classical model demonstrated a slight advantage, followed by the CV quantum model, and finally the DV quantum model. Similarly, in plot (b), classification performance decreases for all models as more noise is injected into the images, however, both the classical and CV quantum models showcased similar performance, outperforming the DV quantum model significantly, albeit, performance for all models is considerably poor for the OrganAMNIST due to its high complexity. Finally, plot (c) corresponding to the results on the BreastMNIST dataset, shows better robustness but high instability for all models, specially for both quantum models, as their F1 score varies significantly as noise increases. Nevertheless, the classical model attained the highest F1 score at the highest noise injection, while the proposed CV quantum model achieved the second highest, and finally the proposed DV quantum model showed the least noise robustness in this case.

\begin{figure}[ht]
\centering

\includegraphics[width=0.3\textwidth]{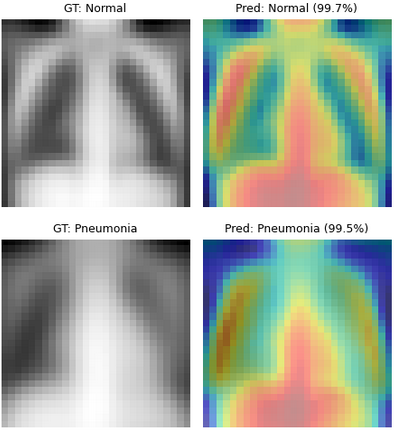}\hspace{0.25cm}
\includegraphics[width=0.3\textwidth]{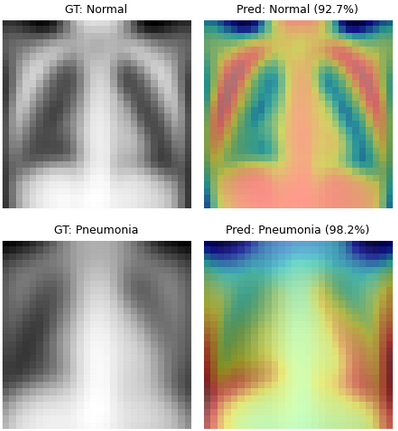}\hspace{0.25cm}
\includegraphics[width=0.3\textwidth]{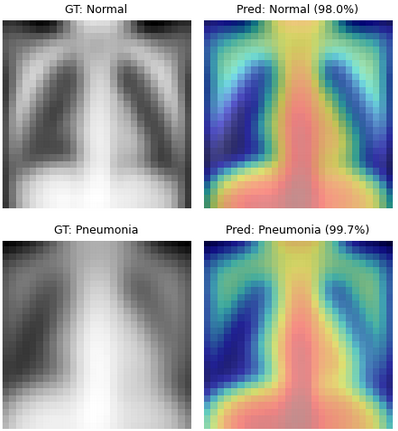}

\vspace{0.3cm}
{\small (a) PneumoniaMNIST heatmaps}

\vspace{0.4cm}

\includegraphics[width=0.3\textwidth]{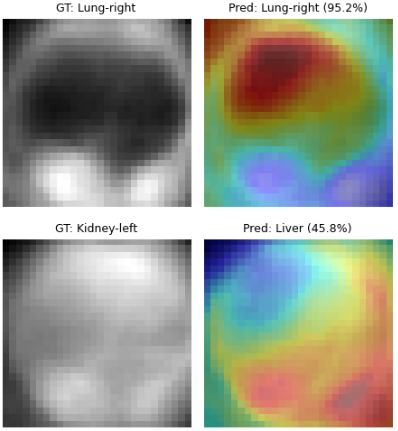}\hspace{0.25cm}
\includegraphics[width=0.3\textwidth]{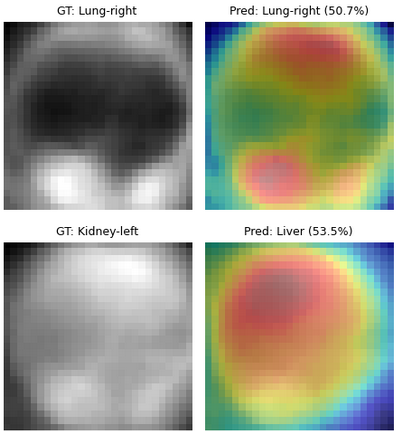}\hspace{0.25cm}
\includegraphics[width=0.3\textwidth]{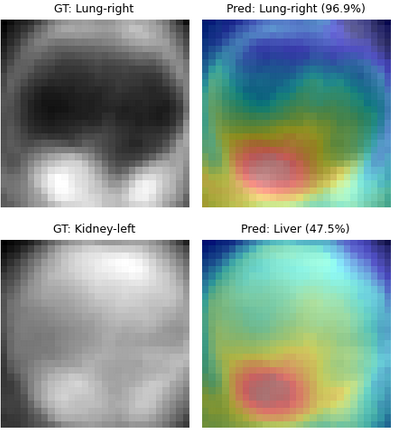}

\vspace{0.3cm}
{\small (b) OrganAMNIST heatmaps}

\vspace{0.4cm}

\includegraphics[width=0.3\textwidth]{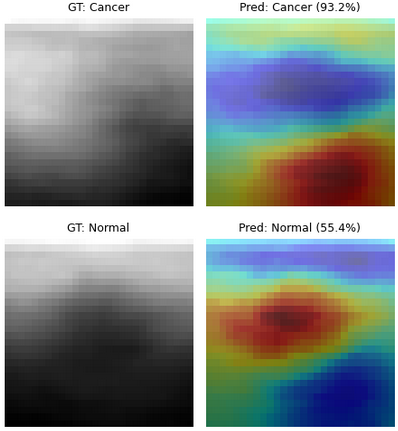}\hspace{0.25cm}
\includegraphics[width=0.3\textwidth]{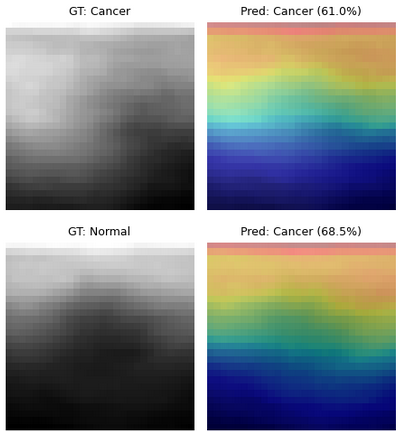}\hspace{0.25cm}
\includegraphics[width=0.3\textwidth]{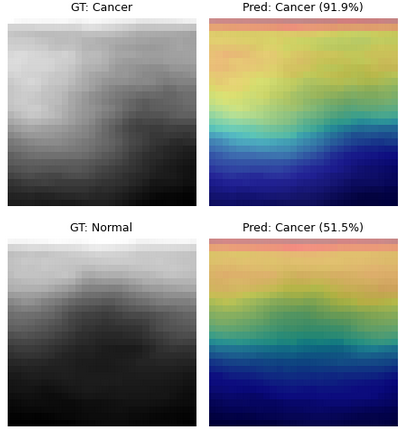}

\vspace{0.3cm}
{\small (c) BreastMNIST heatmaps}

\vspace{0.4cm}

{\footnotesize
\begin{tabular}{ccc}
CV QNN heatmaps & DV QNN heatmaps & Classical NN heatmaps
\end{tabular}
}

\caption{
Comparison of Grad-CAM heatmaps across models and datasets.  
Each row corresponds to a dataset — (a) PneumoniaMNIST, (b) OrganAMNIST, (c) BreastMNIST — and each column compares the Continuous-Variable (CV) QNN, Discrete-Variable (DV) QNN, and classical neural network.  
Red regions indicate the most influential image areas for model predictions.
}
\label{fig:model_heatmaps}
\end{figure}

\textbf{Experiment 5: Decision heatmap comparison on all datasets}\\
To evaluate the potential clinical interpretability of the proposed quantum neural networks, Grad-CAM heatmaps were computed for each model, as shown in Figure \ref{fig:model_heatmaps}. The figure displays reconstructed PCA-encoded images (four components) alongside their corresponding decision heatmaps, highlighting each model's region of interest during prediction. Each $6\times2$ block corresponds to a specific model as indicated by the bottom captions, with the prediction confidence shown above each pair.

For the PneumoniaMNIST samples, clear differences are apparent in the localization patterns. The classical model exhibits more sharply defined \textit{cold} regions, suggesting a focus on the central thoracic structures typical of normal scans. In contrast, both quantum models display broader \textit{warm} regions concentrated over the lungs, particularly in pneumonia cases, indicating a higher sensitivity to areas associated with pathological features. Interestingly, the CV QNN achieves the highest prediction confidence across these examples; however, this should not be interpreted as higher diagnostic accuracy, since all three models correctly classified the samples. Instead, it may indicate a different internal confidence due to the CV model's continuous activation characteristic.

For the OrganAMNIST dataset, model difference become more pronounced. In the ``Lung-right'' case, the CV QNN emphasizes darker areas, potentially corresponding to denser tissue, while the DV QNN distributes its attention more uniformly, with stronger activation in the lower region. The classical model also focuses on the bottom portion, aligning more closely with the CV QNN in spatial attention and confidence (95.2\%). For the ``Kidney-left'' sample, all models misclassify the image as ``Liver'', however, their attention maps differ. The CV QNN and classical model display similar behavior with more localized activations, whereas the DV QNN emphasizes a distinct top-central zone.

In contrast, for the BreastMNIST dataset, the DV QNN and classical model exhibit nearly identical attention distributions, both showing a layered top-to-bottom gradient of relevance. However, the CV QNN displays a more heterogeneous sensitivity pattern, highlighting different regions across samples. This broader activation could reflect a different encoding of image intensity features in continuous-variable space. Notably, this behavior aligns with the CV QNN's higher true-negative performance observed in Experiment 3. These findings suggest that, while quantum and classical models may converge on similar predictive outcomes, their internal mechanisms of attention and sensitivity to spatial features can differ significantly.

\textbf{Experiment 6: Hypothesis Testing}\\
To statistically determine significant difference in classification performance between the implemented models, a hypothesis test through a non-parametric Friedman test, as well as pairwise Wilcoxon signed-rank tests are conducted using the threefold cross-validation F1 score results. The formulated hypothesis to reject or corroborate are the following:
\begin{itemize}
    \item $H_{0}$: There is no statistically significant difference in F1 score between models (Retained if $p\geq0.05$).
    \item $H_{1}$: At least one model differs significantly in F1 score from the others (Accepted if $p<0.05$).
\end{itemize}
The Friedman test is applied to compare the multiple models over the same cross-validation folds. If the null hypothesis was rejected, pairwise signed-rank Wilcoxon test to compare each model is conducted. To control the family-wise error rate due to multiple comparisons, the Bonferroni correction was applied, setting the adjusted level to $\alpha=0.05/3 = 0.0167$. Table \ref{tab:full_stats_datasets} summarizes the statistical outcomes across datasets, reporting each model's mean and standard deviation for all four metrics (accuracy, recall, precision, and F1 score), the chi-square statistic $X^2(2)$ with two degrees of freedom, the Wilcoxon signed-rank statistic $W$, and the corresponding probability values $p$.
\begin{table}[ht]
\centering
\caption{Statistical comparison among Classical, DV, and CV quantum models across the PneumoniaMNIST, OrganAMNIST, and BreastMNIST datasets and the four classification metrics. Values represent mean $\pm$ standard deviation across cross-validation folds. Friedman and Wilcoxon signed-rank tests (Bonferroni-corrected $\alpha = 0.0167$) revealed no statistically significant differences among models ($p > 0.05$).}
\label{tab:full_stats_datasets}

\resizebox{\textwidth}{!}{%
\setlength{\tabcolsep}{2.5pt}
\renewcommand{\arraystretch}{1.75}
\tiny
\begin{tabular}{llcccccccccc}
\hline
\textbf{Dataset} & \textbf{Metric} & \textbf{Classical} & \textbf{DV} & \textbf{CV} & $\boldsymbol{\chi^2(2)}$ & \textbf{p} & $\boldsymbol{W}$ & \textbf{C--DV (p)} & \textbf{C--CV (p)} & \textbf{DV--CV (p)} & $\boldsymbol{\alpha}$ \\
\hline
\multirow{4}{*}{\textbf{PN}} 
 & ACC  & 0.900 $\pm$ 0.008 & 0.899 $\pm$ 0.002 & 0.864 $\pm$ 0.059 & 0.441 & 0.441 & 0.273 & 1.000 & 0.500 & 0.180 & 0.0167 \\
 & P & 0.900 $\pm$ 0.007 & 0.898 $\pm$ 0.002 & 0.870 $\pm$ 0.045 & 0.097 & 0.097 & 0.778 & 1.000 & 0.250 & 0.250 & 0.0167 \\
 & R    & 0.900 $\pm$ 0.008 & 0.899 $\pm$ 0.002 & 0.864 $\pm$ 0.059 & 0.441 & 0.441 & 0.273 & 1.000 & 0.500 & 0.180 & 0.0167 \\
 & F1        & 0.900 $\pm$ 0.008 & 0.899 $\pm$ 0.002 & 0.864 $\pm$ 0.059 & 0.441 & 0.441 & 0.273 & 1.000 & 0.500 & 0.180 & 0.0167 \\
\hline
\multirow{4}{*}{\textbf{ORG}} 
 & ACC  & 0.534 $\pm$ 0.025 & 0.458 $\pm$ 0.030 & 0.549 $\pm$ 0.004 & 0.097 & 0.097 & 0.778 & 0.250 & 1.000 & 0.250 & 0.0167 \\
 & P & 0.476 $\pm$ 0.020 & 0.414 $\pm$ 0.035 & 0.505 $\pm$ 0.001 & 0.0498 & 0.0498 & 1.000 & 0.250 & 0.250 & 0.250 & 0.0167 \\
 & R    & 0.534 $\pm$ 0.025 & 0.458 $\pm$ 0.030 & 0.549 $\pm$ 0.004 & 0.097 & 0.097 & 0.778 & 0.250 & 1.000 & 0.250 & 0.0167 \\
 & F1        & 0.534 $\pm$ 0.025 & 0.458 $\pm$ 0.030 & 0.549 $\pm$ 0.004 & 0.097 & 0.097 & 0.778 & 0.250 & 1.000 & 0.250 & 0.0167 \\
\hline
\multirow{4}{*}{\textbf{BR}} 
 & ACC  & 0.734 $\pm$ 0.009 & 0.725 $\pm$ 0.019 & 0.690 $\pm$ 0.059 & 0.667 & 0.7165 & 0.111 & 0.500 & 0.500 & 0.750 & 0.0167 \\
 & P & 0.649 $\pm$ 0.096 & 0.613 $\pm$ 0.094 & 0.694 $\pm$ 0.030 & 0.667 & 0.7165 & 0.111 & 0.750 & 0.500 & 0.500 & 0.0167 \\
 & R    & 0.734 $\pm$ 0.009 & 0.725 $\pm$ 0.019 & 0.690 $\pm$ 0.059 & 0.667 & 0.7165 & 0.111 & 0.500 & 0.500 & 0.750 & 0.0167 \\
 & F1        & 0.734 $\pm$ 0.09 & 0.725 $\pm$ 0.019 & 0.690 $\pm$ 0.059 & 0.667 & 0.7165 & 0.111 & 0.500 & 0.500 & 0.750 & 0.0167 \\
\hline
\end{tabular}
}
\end{table}

The Friedman test results reveal no statistical significant difference among the three models across any dataset or metric. Because of this, the pairwise signed-rank Wilcoxon tests are conducted to evaluate pairwise comparisons (Classical-DV, Classical-CV, and DV-CV). However, the results further confirm that the differences between the proposed CV quantum, DV quantum, and classical models were not statistically significant, as their probabilities $p$ are larger than the corrected significance threshold $\alpha=0.0167$, thus, indicating a comparable performance across models.

This outcome suggests that, although the attained results fail to demonstrate quantum advantage over their classical counterpart, under the evaluated datasets and current training configuration, the proposed quantum models perform on par to classical methods. In particular, the CV quantum model offers higher representational capacity for structured and imbalanced biomedical data, as observed in the OrganMNIST dataset. 
\section{Discussion}\label{sec:discussion}
This section presents the classification performance of all implemented models on unseen data samples for all datasets to assess clinical implementation feasibility. Table \ref{tab:classification-compilation} summarizes the proposed models' classification performance across all evaluated datasets. For the BreastMNIST dataset, although the classical model attains a marginal advantage over the proposed quantum models (+1.5\%), all models show favorable performance on the test set, achieving similar results in the metrics of accuracy, recall, precision, and F1 score, and equal area under the ROC and PR curves. This uniformity in performance can be attributed to the dataset's relatively low complexity. Yet, despite reduced data dimensionality and a highly reduced set of parameters, each model demonstrates favorable generalization, indicating that the extracted data representations are expressive enough for binary classification. 

In contrast, the increased complexity and higher class diversity of the OrganAMNIST dataset reveal more pronounced performance disparities. The DV quantum model, shows a drop in predictive ability, obtaining 39.15\% accuracy, recall and F1 score, as well as an AUPRC of 37.54\%. Conversely, the CV quantum and classical models show higher robustness under these conditions, attaining a slightly higher accuracy, recall, and F1 score of 45.63\% and 47.37\%, for each of the models respectively. These results suggest that for this configuration, while the DV quantum model may struggle with multiclass generalization, the CV quantum model can maintain moderate performance closer to its classical counterpart thanks to its continuous Hilbert-space representation.

Finally, for the PneumoniaMNIST dataset, all models achieve their highest performance, benefiting from the binary structure and sufficient data availability for feature generalization. Although all models achieve the same performance for AUROC and AUPRC (92\% and 93\%), the DV quantum and classical models slightly outperform their CV quantum counterpart, attaining an accuracy, recall, and F1 score of 85\%. These results across different metrics reinforces that for different types of datasets such as binary low-complexity datasets, high-complexity and large datasets, as well as small imbalanced datasets, classical and quantum neural networks can converge to comparable decision boundaries. 
\begin{table}[ht]
\centering
\caption{Test set classification metrics for CV quantum model, DV quantum model, and classical model across MedMNIST datasets. AUROC and AUPRC results are averaged for the OrganAMNIST dataset. Best results per dataset in bold.}
    \setlength{\tabcolsep}{3.5pt}
    \renewcommand{\arraystretch}{1.75}
\label{tab:classification-compilation}
\tiny
\begin{tabular}{llcccccc}
\hline
\textbf{Model} & \textbf{Dataset} & \textbf{ACC} & \textbf{P} & \textbf{R} & \textbf{F1} & \textbf{AUROC} & \textbf{AUPRC} \\
\hline

\multirow{3}{*}{CV QNN} 
 & BreastMNIST & 0.7564 & 0.7564 & 0.7317 & 0.7564 & 0.73 & \bf 0.86 \\
 & OrganAMNIST     & 0.4563 & 0.4563 & 0.4257 & 0.4563 & 0.8333 & 0.4554 \\
 & PneumoniaMNIST    & 0.8429 & 0.8429 & 0.8437 & 0.8429 & \bf 0.92 & \bf 0.93 \\

\hline
\multirow{3}{*}{DV QNN} 
 & BreastMNIST & 0.7372 & 0.7372 & 0.7662 & 0.7372 & 0.67 & 0.84 \\
 & OrganAMNIST     & 0.3915 & 0.3915 & 0.3714 & 0.3915 & 0.8154 & 0.3754 \\
 & PneumoniaMNIST   & \bf 0.8542 & \bf 0.8542 & 0.8534 & 0.8526 & \bf 0.92 & \bf 0.93 \\

\hline
\multirow{3}{*}{Classical} 
 & BreastMNIST & \bf 0.7628 & \bf 0.7628 & \bf 0.7662 & \bf 0.7628 & \bf 0.74 & \bf 0.86 \\
 & OrganAMNIST     & \bf 0.4737 & \bf 0.4737 & \bf 0.4355 & \bf 0.4737 & \bf 0.8518 & \bf 0.49 \\
 & PneumoniaMNIST    & \bf 0.8542 & \bf 0.8542 & \bf 0.8540 & \bf 0.8542 & \bf 0.92 & \bf 0.93 \\

\hline
\end{tabular}
\end{table}

\section{Conclusion and Future Work}\label{sec:conclusionAndFutureWork}
This research introduces a small-scale CV quantum neural network for biomedical image classification for binary and multiclass classification of the PneumoniaMNIST, BreastMNIST, and OrganAMNIST dataset. The attained results are compared to its DV quantum and classical counterparts. The proposed CV quantum model is a 4-qumode variational quantum circuit of a depth of 2. This circuit encodes pca-encoded reduced input images into quantum states, which are later processed through a series of displacement, squeezing, rotation, and beam splitter Gaussian gates for feature extraction, and a measurement on the position quadrature is conducted to obtain the quantum model's output before integrating it to a fully connected layer that provides the final model prediction. Similarly, the proposed DV quantum neural network follows the same architecture, but with analogous discrete quantum gates, such as rotational, phase, and entanglement gates, which behave akin to the CV quantum model. To assess their performance, classification performance evaluation, noise robustness testing, decision heatmap comparison, and statistical analysis are conducted. The conclusions deriving from the results of these experiments are the following:
\begin{itemize}
    \item The CV and DV quantum models attain comparable classification performance in the training, validation, and test sets in comparison to its classical counterpart across all metrics and datasets (F1 scores of 75\% in BreastMNIST, 45\% in OrganAMNIST, 85\% in PneumoniaMNIST). Furthermore, the CV quantum model shows slightly higher performance than its DV quantum counterpart in multiclass classification (+7\% F1 score in OrganAMNIST), as well as a slight advantage in minority class focus (15\% TN samples of BreastMNIST).
    \item The decision heatmaps of the proposed quantum models show similar confidence as their classical counterpart. Specifically, the proposed CV QNN showcases a more comprehensible heatmap in the BreastMNIST dataset, highlighting specific areas, rather than showing layer-like significance interpretation as shown by the DV quantum and classical models. These results suggest better output interpretability due to the increase of Hilbert space data representation, aiding to its potential in clinical implementation.
    \item Noise robustness testing shows high robustness for the CV QNN, as it demonstrates closer classification stability to its classical counterpart than the DV QNN over a range of [0.1, 1.0] of Gaussian noise.
    \item Statistical analysis and hypothesis testing is conducted by comparing the threefold cross-validation performance of the three models, showing no significant difference between all of them in the Friedman test. Additional statistical testing through pairwise signed-rank Wilcoxon tests further corroborates these results. Comparing each pair of implemented models between them, demonstrated no significant difference in their classification performance, showcasing the potential that data encoding and small-scale CV and DV quantum circuits offer to perform on par to classical counterparts.
\end{itemize}
In conclusion, the proposed CV QNN offers comparable performance to its classical and DV quantum counterparts, while also showcasing higher noise robustness, minority class focus, and multiclass classification advantages. This work explores the potential that CV quantum computing offers in biomedical imaging thanks to its increased Hilbert space dimensions, its capability to work with continuous data, nonlinearity encoding, as well as the tradeoff offered by Gaussian gates for feature extraction. Together, these findings highlight the relevance of contributions in the CV quantum machine learning field, as well as promise in clinical practice through quantum technologies.

For following steps, additional data preparation processes can be explored to increase dimensionality reduction and maximize data feature representation. Additionally, more complex datasets can be tested to further assess the proposed models' effectiveness in medical imaging diagnostic tasks. Model architecture, depth, and complexity can also be further developed, introducing non Gaussian gates, increased qumodes/qubits, as well as trainable parameters for higher feature capturing.

\section*{Acknowledgments}
We gratefully acknowledge the support of the Secretaría de Ciencia, Humanidades, Tecnología e Innovación (Secihti), and the Instituto Politécnico Nacional (IPN), which made this research possible.
\section*{Declarations}
\begin{itemize}
    \item \textbf{Consent for publication:} All of the authors reviewed and approved the final manuscript for publication.
    \item \textbf{Data availability:} The MedMNIST dataset used to conduct this research is publicly available. It can be accessed from its official website: \texttt{https://medmnist.com/} as well as from their Github page \texttt{https://github.com/MedMNIST/MedMNIST}, and publication \href{https://arxiv.org/abs/2110.14795}{\texttt{10.48550/arXiv.2110.14795}}.
    \item \textbf{Code availability:} The code used to conduct this research is protected under the MIT license and is available in the following GitHub repository: \texttt{PENDING}.
\end{itemize}

\providecommand{\newblock}{}

\end{document}